\documentclass{aa}  
\usepackage{graphicx}
\usepackage{txfonts}
\usepackage[breaklinks=true,colorlinks=true,urlcolor=blue,citecolor=blue,linkcolor=blue]{hyperref} 
\bibpunct{(}{)}{;}{a}{}{,} 

\newcommand{\be}{\begin{eqnarray}}
\newcommand{\ee}{\end{eqnarray}}
\def\He{\ion{He}{}}
\def\HeI{\ion{He}{I}}
\def\HeII{\ion{He}{II}}
\def\HeIII{\ion{He}{III}}

\def\SiIV{\ion{Si}{IV}}
\def\MgII{\ion{Mg}{II}}

\def\relpop{\ensuremath{n(2\mathrm{s}\,^3\mathrm{S})/n_\He }}
\def\Heline{\HeI~1083~nm}
\def\nne{\ensuremath{n_\mathrm{e}}}
\def\nH{\ensuremath{n_\mathrm{H}}}
\def\Ha{\mbox{H\hspace{0.2ex}$\alpha$}}

\def\Hecont{ \HeI\ ground state continuum}

\def\dd{\mathrm{d}}

\def\Hi{\ion{H}{I}}

\def\ne{\ensuremath{n_\mathrm{e}}}

\def\bifrost{{\it Bifrost}}
\def\multitd{{\it Multi3d}}
\def\radyn{{\it RADYN}}

\def\figspath{.}

\begin{document} 

   \title{The cause of spatial structure in solar \Heline\  multiplet images}

  \author{Jorrit Leenaarts\inst{\ref{Stockholm}}
          \and 
     Thomas Golding\inst{\ref{Oslo}}
          \and 
        Mats Carlsson\inst{\ref{Oslo}}
        \and
        Tine Libbrecht\inst{\ref{Stockholm}}
        \and
         Jayant Joshi\inst{\ref{Stockholm}}
        }
          
   \institute{Institute for Solar Physics, Department of Astronomy, Stockholm University,
AlbaNova University Centre, SE-106 91 Stockholm Sweden, \email{jorrit.leenaarts@astro.su.se,tine.libbrecht@astro.su.se,jayant.joshi@astro.su.se,}\label{Stockholm}
\and
 Institute of Theoretical Astrophysics, University of Oslo, P.O. Box 1029 Blindern, N--0315 Oslo, Norway
 \email{mats.carlsson@astro.uio.no,thomas.golding@astro.uio.no}\label{Oslo}}

   \date{Received ; accepted }

 
  \abstract
   {The \Heline\ is a powerful diagnostic for inferring properties of the upper solar chromosphere, in particular for the magnetic field. The basic formation of the line in one-dimensional models is well understood, but the influence of the complex three-dimensional structure of the chromosphere and corona has however never been investigated. This structure must play an essential role because images taken in \Heline\ show structures with widths down to 100~km.}
  {To understand the effect of the three-dimensional temperature and density structure in the solar atmosphere on the formation of the \Heline\ line.}
   {We solve the non-LTE radiative transfer problem assuming statistical equilibrium for a simple 9-level helium atom that nevertheless captures all essential physics. As a model atmosphere we use a snapshot from a 3D radiation-MHD simulation computed with the \bifrost\ code. Ionising radiation from the corona is self-consistently taken into account.}
   {The emergent intensity in the \Heline\ is set by the source function and the opacity in the upper chromosphere. The former is dominated by scattering of photospheric radiation and does not vary much with spatial location. The latter is determined by the photonionisation rate in the \Hecont, as well as the electron density in the chromosphere. The spatial variation of the flux of ionising radiation is caused by the spatially-structured emissivity of the ionising photons from material at $T \approx100$~kK in the transition region. The hotter coronal material produces more ionising photons, but the resulting radiation field is smooth and does not lead to small-scale variation of the UV flux. The corrugation of the transition region further increases the spatial variation of the amount of UV radiation in the chromosphere. Finally we find that variations in the chromospheric electron density also cause strong variation in \Heline\ opacity. {We compare our findings to observations using SST, IRIS and SDO/AIA data.} }
   {}


   \maketitle
%

\section{Introduction}

The \Heline\ multiplet arises from the three radiative transitions between the 1s2s\,$^3\mathrm{S}$ and 1s2p\,$^3\mathrm{P}$ triplet terms of neutral helium. In the solar spectrum it appears as two absorption lines at 1082.91~nm and 1083.03~nm (in air). Because of the large energy difference between the lower level of the multiplet and the singlet ground state (19.8~eV), the population of the lower level cannot be caused by electron collisions at typical chromospheric temperatures of $\sim10$~kK. This led 
\citet{1939ApJ....89..673G} 
to propose that photoionisation from the \HeI\ ground state continuum ($\lambda<50.4$~nm) {follow by spontaneous recombination (often called PR mechanism)} is the driver of triplet population. Observational evidence from solar observations is the general correlation of X-ray emission and \Heline\ absorption
\citep{1983SoPh...87...47K}.
Observations of weakly active dwarfs, giants and supergiant stars of spectral type F7 or later show a correlation between X-ray flux and \Heline\ equivalent width
\citep{1986ApJ...304..365Z}. 
The same relation has been established for active giants, but interestingly, not for active dwarfs
\citep{2008A&A...488..715S}.

The observational evidence has been supported by many calculations of increasing sophistication 
\citep[e.g.][]{1975ApJ...199L..63Z,1976SoPh...49..315L,1988SvA....32..542P,1994isp..book...35A,1997ApJ...489..375A,%
2008ApJ...677..742C}, 
but all either in a plane-parallel slab geometry or using one-dimensional plane-parallel atmospheres without an actual corona.
Consequently the coronal radiation field in these calculations had to be prescribed, for which typically the data described in
\citet{1991JATP...53.1005T} 
has been used. All models reproduced the observed correlation between the solar EUV flux and the \Heline\ equivalent width
\citep[EW, see][]{1991JATP...53.1005T}.
{An interesting feature of the PR mechanism is that the 30.4~nm of \HeII\ line is the strongest line in the \Hecont, so that the formation of the \Heline\  is influenced by the resonance lines of \HeII.}

\citet{1976SoPh...49..315L}
demonstrated that the relative population density of the triplet system also depends on the electron density. By considering the dominant pathways that populate and depopulate the triplet system they found that the relative population of the triplet ground state \relpop\ is approximately given by
\be
\frac{n(2\mathrm{s}\, ^3\mathrm{S})}{n_\He} = \frac{\alpha_\mathrm{t}}
{\left( 1 + \ne \alpha_\mathrm{s} / \phi_\mathrm{s}  \right) \left(  \phi_\mathrm{t} / \ne + q_\mathrm{ts} \right) },  \label{eq:relpop}
\ee
with $\alpha_\mathrm{t}$ the photorecombination rate coefficient per electron into the triplet system, $\ne$ the electron density, $\alpha_\mathrm{s}$ the photorecomination rate coefficient per electron into the singlet system, $\phi_\mathrm{s}$ and $ \phi_\mathrm{t}$ the photoionization rate coefficients for the 1s$^2$\,$^1\mathrm{S}$ and 1s2s\,$^3\mathrm{S}$ levels, and $q_\mathrm{ts}$ the electron collision rate coefficient per electron from the 1s2s\,$^3$S level into the 1s2p\,$^1$P and 1s$^2$\,$^1$S singlet levels. The quantities $\alpha_\mathrm{t}$ and $\alpha_\mathrm{s}$ are approximately constant (exactly constant if stimulated recombination is ignored), $\phi_\mathrm{s}$ depends on the radiation field blueward of 50.4~nm, $\alpha_\mathrm{t}$ depends on the radiation field blueward of 259~nm, and $q_\mathrm{ts}$ depends on temperature. 

The relative population is an increasing function of electron density for typical values of the upper chromospheric electron density ($10^{15}$--$10^{18}$ m$^{-3}$) and temperature (10~kK) at fixed coronal UV illumination. We thus also expect variations in the \Heline\ strength caused by variations in the mass density and ionisation degree in the chromosphere. 

{An alternative mechanism for creating \Heline\ opacity is direct collisional excitation from the \HeI\ ground state to  the lower level of the line. Owing to the large energy gap between the two levels this mechanism is inefficient in the chromosphere, but can act whenever the temperature is above 20,000~K, i.e, in the lower transition region. 
\citep[e.g.][]{1973ApJ...186.1043M}.
At those temperatures one expects little \HeI\ based on statistical equilibrium calculations, but it might be present due to slow ionization and recombination during rapidly changing thermodynamic properties and the a drift of \HeI\ atoms into areas with higher electron temperature in the presence of strong temperature gradients and a long mean time between collisions.
Several authors have proposed such mechanisms to explain the intensities of the resonance lines  of helium, such as \HeI\ 53.7~nm, \HeI\ 58.4~nm, \HeII\  30.4~nm, and HeII\ 25.6~nm.
\citep[][]{1973ApJ...186.1043M,1975MNRAS.170..429J,2002MNRAS.337..666S,2003MNRAS.341..143S,2004ApJ...606.1239P}.
Such effects also influence the formation of the \Heline\ in 1D radiation-hydrodynamics simulations
\citep{2014ApJ...784...30G}.
}

The one-dimensional nature of the modelling efforts so far has meant that they are unable to explain the spatial structure of the \Heline\ equivalent width, besides the general conclusion that large EW corresponds to larger impinging EUV flux{ and/or larger chromospheric density.}

High-resolution coronal images 
\citep[e.g.,][]{2013Natur.493..501C}
show spatial structure down to $0\farcs2$, indicating coronal temperature and/or density variations at least down to that scale, while the actual emitted spectrum below 50.4~nm strongly depends on temperature. This leads to a large variation in space of the emissivity of ionising radiation, and this should lead to a spatial variation of the ionising radiation impinging on the chromosphere. The transition region between the chromosphere and the corona is highly corrugated which will introduce additional differences in impinging flux as parts of the chromosphere can shield other parts from the coronal radiation. 

Three-dimensional numerical simulations 
\citep[e.g.,][]{2016A&A...585A...4C}
as well as semi-empirical 1D models of the solar chromosphere
\citep[e.g.,][]{2006ApJ...639..441F}
indicate that considerable density variations occur at the top of the chromosphere. 

One thus expects strong spatial variation of the strength of the  \Heline. This is indeed the case: high-resolution observations show spatial structure down to a fraction of an arcsecond
\citep[e.g.][]{2012ApJ...750L..25J,2013ApJ...768..111S,2015SoPh..290.1607S}. 

In this paper we investigate the causes of the spatial variation of the \Heline\ strength with help of numerical simulations of the solar atmosphere, 3D non-LTE radiative transfer calculations, and observations taken with the Swedish 1-m Solar Telescope (SST) and the Interface Region Imaging Spectrograph (IRIS). Our aim is to investigate in detail the origin of the variation in ionising UV radiation as well as the influence of the corrugation of the transition region and the chromospheric electron density variation on  \Heline\ images.

In Section~\ref{sec:method} we describe our numerical simulation and the subsequent radiative transfer modelling. Section~\ref{sec:observations} describes our observations. We describe the relation between coronal UV radiation, chromospheric electron density and \Heline\ absorption in Section~\ref{sec:results}. Section~\ref{sec:test} contains a brief observational test of our theoretical finding that \Heline\ absorption should correlate with \SiIV\ emission, and we finish with a discussion in Section~\ref{sec:conclusions}.

\section{Numerical Method} \label{sec:method}

\paragraph{Radiation-MHD simulation} \label{sec:bifrost}

As a model atmosphere we use a snapshot from a three-dimensional radiation-MHD simulation computed with the \bifrost\ code
\citep{2011A&A...531A.154G}. 
This code solves the resistive-MHD equations together with heat conduction and non-LTE radiative losses on a Cartesian grid. The particular simulation that we use here is a 3D extension of the 2D simulations of
\citet{2016ApJ...817..125G}. 
It uses an equation-of-state (EOS) that takes the non-equilibrium ionisation of hydrogen and helium into account. Non-equilibrium ionisation has a strong effect on the temperature and density structure in the chromosphere and transition region, and is thus required to obtain a realistic model of the upper chromosphere
\citep{2002ApJ...572..626C,2007A&A...473..625L,2014ApJ...784...30G,2016ApJ...817..125G}.

The inclusion of the Lyman continuum and the Lyman alpha line in \bifrost\ is important because the hydrogen Lyman continuum is a source of ionising photons in the He I continuum. Treating these these transitions in detailed balance, as was done in earlier models
\citep[such as those described in][]{2016A&A...585A...4C}
leads to an erroneous amount of Lyman continuum photons in the subsequent radiative transfer step.

The simulation was run on a grid of $504 \times 504 \times 496$ points, with a horizontal domain size of $ 24 \times 24$~Mm$^2$. In the vertical direction the domain spans from 2.4 Mm below the average height of $<\tau_{500}>=1$ to 12.4~Mm above it. Besides the different EOS the run has an identical setup to the publicly available simulation described in
\citet{2016A&A...585A...4C}. 
In short, the simulation was run using an LTE EOS for 1,750~s after an initial smooth bipolar magnetic field was introduced into a previously relaxed hydrodynamical simulation. This removed transients caused by the introduction of the magnetic field. Then the EOS was changed from LTE to non-equilibrium ionisation of hydrogen and helium, and the simulation was run for an additional 2,000~s. We took the snapshot 780~s after the non-equilibrium EOS was switched on. We refer the reader to
\citet{2011A&A...531A.154G},
\citet{2016ApJ...817..125G},
and
\citet{2016A&A...585A...4C}
for further details of the simulation.

\paragraph{Model atmosphere}

The \bifrost\ snapshot was reduced in size to a grid of $252\times252\times496$ by removing every second column in the two horizontal directions to reduce the computational cost of the radiative transfer computation.

\paragraph{Model atom}

\begin{figure}
   \centering
   \includegraphics[width=\hsize]{\figspath/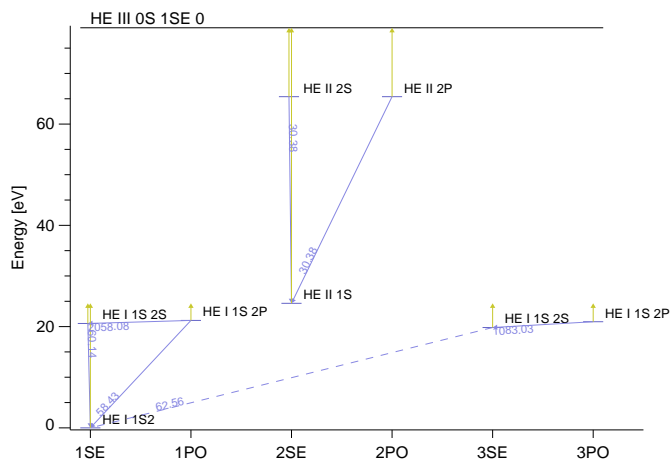}
      \caption{Term diagram of our nine-level helium model atom. Solid blue lines indicate allowed radiative bound-bound transitions, dashed lines forbidden bound-bound transitions and yellowish lines radiative bound-free transitions.}
              \label{fig:termdiagram}
    \end{figure}

We use the same helium model atom as 
\citet{2002ApJ...572..626C}.
By collapsing the sublevels in each term to a collective level this atom has only nine states. In this way the atom includes the 1s$^2$, 1s2s, and 1s2p terms of the singlet system and the 1s2s and 1s2p terms of the triplet system of \HeI, the 1s, 2s and 2p terms of \HeII\ and a single state for \HeIII. This model captures all pertinent processes despite its small size: it includes the bound-free transitions between the ground states that give rise to the ionisation edges at 22.8~nm and 50.4~nm, the \HeII\ 30.4 nm line, and a representative  \Heline\ line. The latter does not have the three components (from the upper levels 2\,$^3\mathrm{P}_2$, 2\,$^3\mathrm{P}_1$, and 2\,$^3\mathrm{P}_0$ to the lower level 2\,$^3\mathrm{S}_1$) as in reality, but instead consist of a single transition at 1083.33~nm in vacuum with an oscillator strength $f=0.539$. A term diagram of our model atom is shown in Fig.~\ref{fig:termdiagram}.

\paragraph{Radiative transfer}

The radiative transfer calculations are done with the \multitd\ code
\citep{2009ASPC..415...87L}.
It solves the problem of non-LTE radiative transfer in full 3D using the formalism of 
\citet{1991A&A...245..171R,1992A&A...262..209R}.
The code is parallellized using MPI and employs domain decomposition in the three spatial dimensions as well as in  frequency. The formal solution is computed on short characteristics using monotonic cubic Hermite polynomials 
\citep{2003ASPC..288....3A,2013A&A...549A.126I}
to interpolate the source function and the extinction coefficient. Interpolation from grid points to cell interfaces is done using cubic convolution modified to avoid overshooting artefacts. As angle quadrature we use the A4 set from 
\citet{carlson1963}, which contains 24 angles.
We compute the background opacities, and background scattering probabilities in the photosphere and chromosphere for all elements except hydrogen and helium in LTE using the Uppsala opacity package 
\citep{gustafsson1973}.
The hydrogen opacities and scattering probabilities are based on the non-equilibrium hydrogen populations from the \bifrost\ simulation (see Sec.~\ref{sec:bifrost}) instead. Note that the background opacities include a scattering term assuming coherent scattering instead of complete redistribution. This assumption influences the Lyman-continuum intensities, which in turn has an effect on the photoionisation in the \Hecont. Ideally one would like to run both hydrogen and helium simultaneously in full non-LTE, but this is currently not possible with \multitd.

\paragraph{Applicability of statistical equilibrium for helium}

In %
\citet{2014ApJ...784...30G,2016ApJ...817..125G}
we showed that the ionisation of helium is out of statistical equilibrium. Comparisons of one-dimensional calculations assuming SE and NE using the \radyn\ code
\citep[e.g.,][]{1992ApJ...397L..59C}
indicate that significant differences in the \Heline\ profiles can occur between SE and NE calculations
\citep[see Fig.~{9} of][]{2014ApJ...784...30G}.
We therefore expect that the current SE calculations will be substantially different from a more physically correct NE calculation. The latter is however currently beyond our means. We focus here on the spatial structure of \Heline\ images caused by the spatial variation in the irradiation from the corona and variations in the chromospheric electron density, which will be qualitatively reproduced by our SE approach. We do not attempt a detailed analysis of the line profiles because they will be quantitatively affected by NE effects.

\paragraph{Coronal radiation}
We include radiation emitted in the corona by adding a coronal emissivity:
\be
\label{eq:coronal_emm}
\psi_\nu = \Lambda_\nu(T) \nne \nH,
\ee
where $\psi_\nu$ is the emissivity in units of power per solid angle per frequency per volume,  $\nne$ and $\nH$ are the electron density and hydrogen density and $\Lambda_\nu(T)$ is the emissivity per hydrogen atom per electron. The quantity $\Lambda_\nu(T)$ was computed using the CHIANTI package version 7.1
\citep{1997A&AS..125..149D,2013ApJ...763...86L}. 
For a grid of 121 temperatures, logarithmically spaced between 1 kK and 10 MK, we computed the radiative losses assuming coronal equilibrium ionisation balance for all lines in the CHIANTI database using the abundance given in the file \texttt{sun\_coronal.abund} and the ion fractions given in the file \texttt{chianti.ioneq}. We removed all lines from hydrogen and helium because they are accounted for already in \multitd: helium as the active non-LTE atom and hydrogen as a background opacity source. The resulting emissivity was binned into 0.1~nm wide bins such that the frequency integral yields the total radiative losses at that temperature. This table was remapped to the coarser frequency grid used in \multitd\ while conserving the frequency integral.

The ionisation state in the time-varying corona is not in equilibrium
\citep[e.g.,][]{1979ApJS...40..793J,1993ApJ...402..741H,2013SSRv..178..271B}, 
and this will have an effect on the coronal emissivities. While it is possible to run \bifrost\ with non-equilibrium coronal ionisation 
\citep{2013AJ....145...72O},
it is computationally prohibitively expensive to include all elements and atomic levels present in the CHIANTI line list. We therefore opted for the simpler and faster coronal equilibrium approximation.

\section{Observations} \label{sec:observations}

The observations were taken on 2015-07-31 from 08:46:36~UT to 08:50:18~UT with the TRIPPEL spectrograph
\citep{2011A&A...535A..14K}
 at the Swedish 1-m Solar Telescope 
\citep[SST, ][]{2003SPIE.4853..341S} 
on La Palma. We obtained raster scans of active region NOAA 12393 at a viewing angle $\mu=0.76$, recorded in reasonable but not excellent seeing conditions. The raster scan contains 300 slit positions with a step size of $0\farcs1$ and a slit width of $0\farcs11$. The field-of-view is $30\arcsec \times 34\farcs2$. It took 222~s to record the full scan.

The SST was used for the first time to observe in the infrared, employing an OWL-camera (OWL SW1.7 CL-640). The recorded spectral region ranges from 1081.84 nm to 1083.49 nm featuring the \Heline\ and \ion{Si}{I} 1082.7 nm spectral lines. The estimated spectral resolution of TRIPPEL is $\lambda / \delta \lambda \approx 150,000$ as determined by comparing with a high resolution reference spectrum
 \citep{2011A&A...535A..14K}.

The sensor of the OWL-camera has a size of $640\times 512$ pixels with a spectral dispersion of 2.6 pm/pixel and a spatial scale of $0\farcs067$~pixel$^{-1}$. The spectra were rebinned in the spatial direction to a scale of 0.1 arcsec/pixel to create square pixels in the raster scan.  We took three acquisitions in each slit position with an exposure time of 100~ms each, with a 0.74~s cadence per slit position. For each slit position we used the acquisition that had maximum continuum contrast.

We reduced the spectra using the TRIPPEL pipeline 
 \citep{2009A&A...507..417P,2011A&A...535A..14K}.
The reduction steps consist of flat field and dark frame corrections, geometrical corrections for smile and keystone and a wavelength calibration. The spectra were corrected for stray-light and hot pixels were removed from the spectra. 

The SST observations were accompanied by observations with the Interface Region Imaging Spectrograph 
 \citep[IRIS,][]{2014SoPh..289.2733D}.
The IRIS continuum channel at 283.2 nm was used for alignment with the SST raster scan continuum image: First, the IRIS slit-jaw images were interpolated to the scale of the SST raster scan, rotated and aligned with the slit jaw image. Second, to correct for  higher-order shifts due to seeing, we applied a destretching routine from the CRISPRED pipeline 
\citep{2015A&A...573A..40D}
to the SST raster scan with the seeing-free IRIS 283.2~nm continuum image taken half-way during the SST scan as an anchor.  As a result of this correction, we are confident that the alignment has accuracy of $\sim 0\farcs2$  which is below the IRIS resolution of $ \sim 0\farcs35$. 

{We also make use of co-spatial and co-temporal data taken with the Atmospheric Imaging Assembly 
\citep[AIA,][]{2012SoPh..275...17L}
on board the Solar Dynamics Observatory 
\citep[SDO,][]{2012SoPh..275....3P}.
These data were co-aligned with the IRIS and SST data to within an accuracy of one AIA pixel,}

In this paper we compare the SST raster scan with IRIS \SiIV\ 140 nm slit-jaw images (from now on \SiIV\ SJI) {and the AIA data}. Therefore, we created an artificial raster scan with the same slit positions and cadence as the SST raster scan from the  \SiIV\ SJI images. Each column in this artificial raster scan equals the column in the \SiIV\ SJI which is the nearest neighbour in time to the SST acquisition time of the column. Due to the big difference in cadence (0.74 s per slit position in the SST and 17 s cadence for the \SiIV\ SJI), only 14 slit-jaw images are used to compose this artificial raster scan.  {Artificial rasters of the AIA data (12~s cadence) where created in the same way as was done for the IRIS data.}

\section{Results} \label{sec:results}

\subsection{Coronal sources of ionising radiation in the \Hecont} \label{sec:coronal_sources}

  \begin{figure}
   \centering
   \includegraphics[width=\hsize]{\figspath/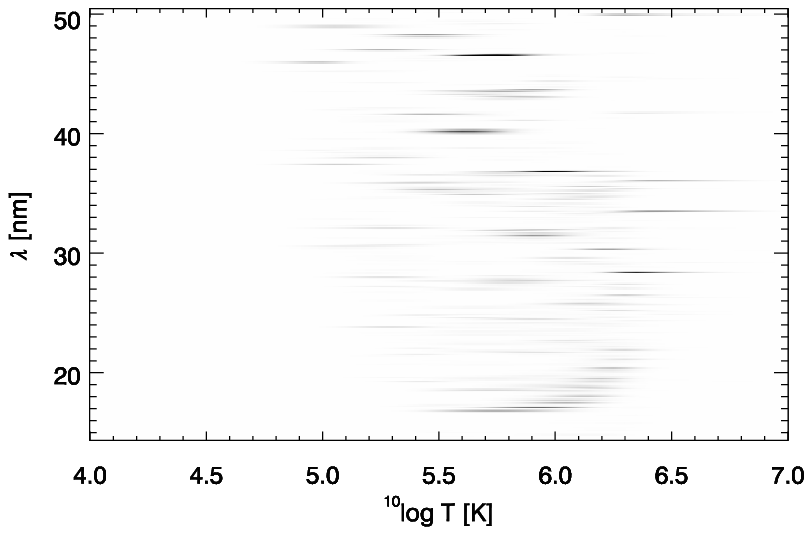}
    \includegraphics[width=\hsize]{\figspath/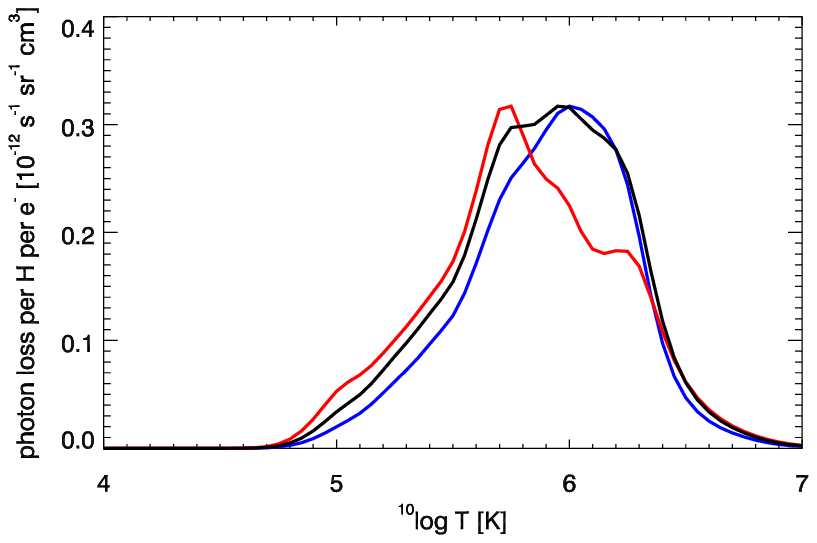}
      \caption{Top: Coronal radiation losses per hydrogen atom per electron  ($\Lambda_\nu(T)$ in Eq.~\ref{eq:coronal_emm}) as function of temperature and wavelength. The scale is inverted, so black means large losses. 
Bottom:{  $\Lambda_\mathrm{E}$, the frequency-integrated energy losses per hydrogen atom per electron (blue, arbitrary units); 
$\Lambda_\mathrm{ph}$, the frequency-integrated photon losses per hydrogen atom per electron (black, scale on the left); }
$\Lambda_\mathrm{csw}$, the frequency-integrated, \HeI\ ground-state continuum cross section weighted, photon losses per hydrogen atom per electron (red, arbitrary units).
        \label{fig:photon_prod}}
    \end{figure}

Our 3D Bifrost model includes the lower corona with variations in temperature and density, so it should mimic the observed behaviour. We investigate systematically how the EUV radiation is produced. 

{As a first check we compared the average UV flux in our model to the values in the {''low activity model'' of 
\citet{1991JATP...53.1005T}.}
{Newer models are available \citep[see][]{2004AdSpR..34.1736T}, but the 1991 model is sufficiently accure for our purposes.}
We integrated the average vertically emerging intensity in our model from 14.3~nm to 50.4~nm, the range over which we include  absorption and emission in the \Hecont, and found a value of 48~W~m$^{-2}$~ster$^{-1}$. This is $\sim60$\% higher than the corresponding value computed from
\citet{1991JATP...53.1005T}.
We thus conclude that our model produces a UV flux in the  \Hecont\ comparable to observations. }

{Note that we do not include absorption in the \HeI\ continuum below 14.3~nm, for which Tobiska gives 3.7~W~m$^{-2}$~ster$^{-1}$, so the error is small. }
{Also note that the Tobiska results are consistent with newer observations with AIA/EVE
\citep{2012SoPh..275..115W}.}

In Fig.~\ref{fig:photon_prod} we show $\Lambda_\nu(T)$ in Eq.~\ref{eq:coronal_emm} for $\lambda<51$~nm. The strongest emission feature is a \ion{Ne}{VII} resonance line at 46.522~nm at $\log_{10} T=5.7$. The blue curve in the bottom panel shows  $\Lambda_E = \int_{\nu_\mathrm{min}}^{\nu_\mathrm{max}} \Lambda_\nu  \, \dd\nu$, the losses  integrated over the frequency range in  which we include \Hecont\ opacity in our model atom. This distribution peaks at $\log_{10} T=6$.

In black we show the frequency-integrated photon losses, to take into account that each emitted photon can only ionise one \HeI\ atom (we omit the integration bounds for brevity):
\be
\Lambda_\mathrm{ph}=\int \frac{\Lambda_\nu(T)}{h \nu}  \, \dd\nu.
\ee
The photon-energy weighting increases the relative contribution of lower-temperature gas to the \HeI\ photoionisation because of the weighting with the inverse frequency.  Finally we computed the coronal photon losses as the \Hecont\ experiences it, by also weighting with the photoionisiaton cross section:
\be
\Lambda_\mathrm{csw}=\int \sigma_\nu \frac{ \Lambda_\nu(T)}{h \nu}  \, \dd\nu, 
\ee
shown in red in the bottom panel of Fig.~\ref{fig:photon_prod}. The peak of the distribution then shifts to $^{10}\!\log T=5.7$~K, owing to the \ion{Ne}{VII} line. In summary, the \Hecont\ is sensitive to radiation emitted at a wide range of temperatures, with peak sensitivity at 500~kK plasma.

  \begin{figure}
   \centering
   \includegraphics[width=\hsize]{\figspath/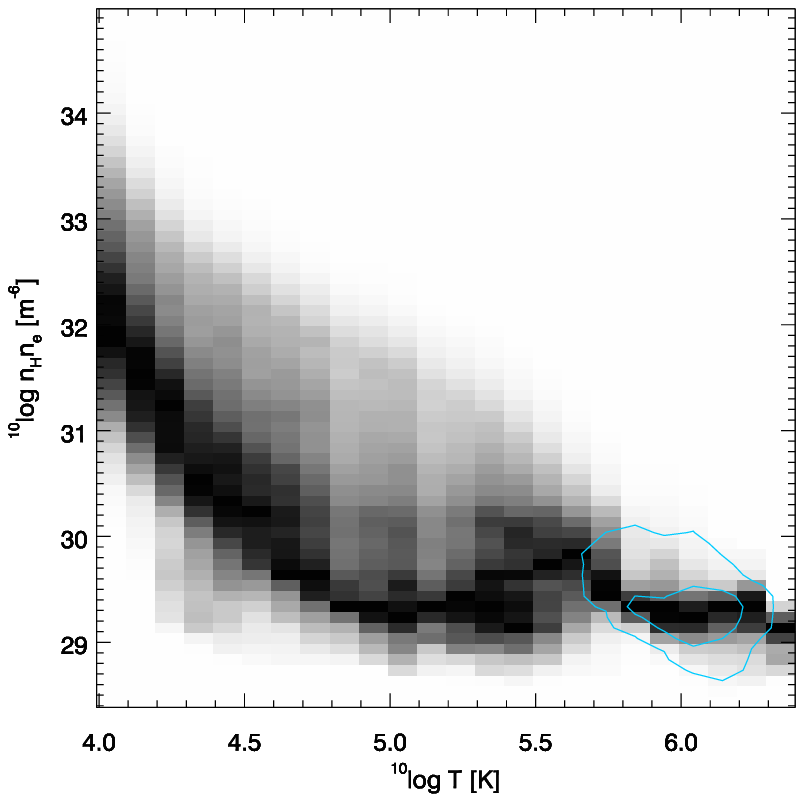}
    \includegraphics[width=\hsize]{\figspath/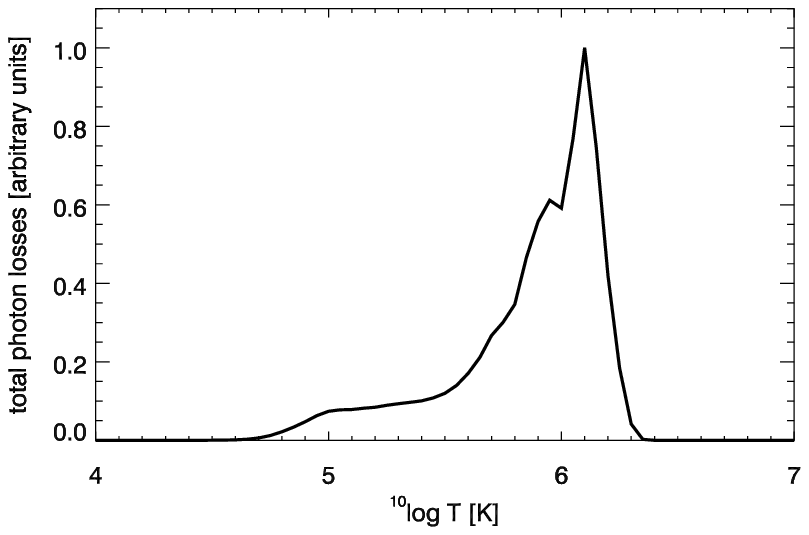}
      \caption{Top: Scaled joint probability density functions (JPDF) of $\nH\nne$ versus $T$ in the chromosphere and corona of our simulation snapshot. The inner blue contour includes 50\% of all pixels, the outer contour 75\%. Each column in the panels is scaled to maximum contrast to increase visibility. Bottom: volume-integrated, frequency integrated and cross-section weighthed photon losses as function of temperature in our simulation snapshot (i.e.,  $\int_V L_\mathrm{csw}  \, \dd V$ in Eq.~\ref{eq:lcsw}). 
      \label{fig:dens_dis}}
    \end{figure}

We now proceed to investigate the effect of the density variations in the transition region and corona of the atmosphere model. Figure~\ref{fig:dens_dis} shows the distribution of $\nH\nne$ versus the temperature in the chromosphere and corona of our simulation snapshot. The majority of the volume is filled with gas at coronal temperatures, but this gas has a relatively low density. The colder material at transition-region temperatures occupies a much smaller volume, but has a much higher density.
Combining all ingredients we finally arrive at the bottom panel of Fig.~\ref{fig:dens_dis}, which shows 
\be
\int_V \Lambda_\mathrm{csw} \nH \nne  \, \dd V = \int_V L_\mathrm{csw}  \, \dd V \label{eq:lcsw}
\ee
the cross-section weighted total production of ionising photons in our simulation volume $V$ as function of temperature. The majority of the photons are produced by the tenuous coronal gas but there is a significant tail toward transition region temperatures.

  \begin{figure*}
   \centering
   \includegraphics[width=17cm]{\figspath/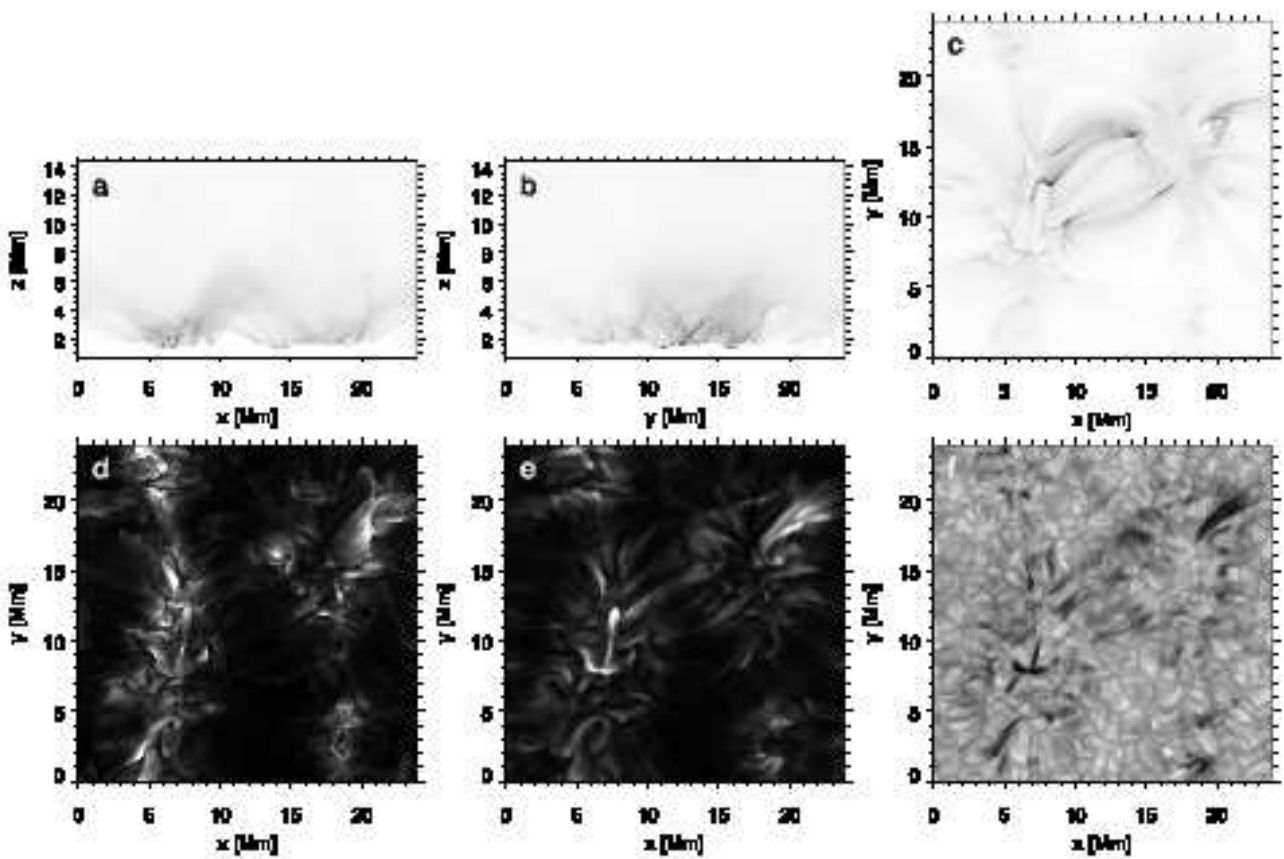}
      \caption{Spatial relation between coronal emissivity and opacity in \Heline. Top row: frequency integrated and cross-section weighted photon losses $L_\mathrm{csw}$, integrated along the $y$-axis (a), the $x$-axis (b) and the $z$-axis (c), on an inverted brightness scale. Panel (d): the resulting angle-averaged radiation field at the largest height in each column where $T=15$ kK, showing the spatial variation of the ionising radiation impinging on the chromosphere. Panel (e): total depth-integrated column density of the lower level of the \Heline. Panel (f): vertically emergent intensity at the nominal line core of the \Heline. {This figure has an accompanying movie in the online material. The movie shows 2D cross-cuts through the simulation domain, with $\Lambda_\mathrm{csw}$ in red and the \Heline\ lower level population in blue} 
      \label{fig:emm_j_n_corr}}
    \end{figure*}

In Fig.~\ref{fig:emm_j_n_corr} we show the effect of the spatial variation of $L_\mathrm{csw}$. Panels a--c show $L_\mathrm{csw}$ integrated along the $x$,$y$, and $z$-axis, respectively. The distribution along 2D cross cuts are given in the accompanying animation. The emission (red colors in the animation) is highly structured, and peaks in a thin sheath just above the chromosphere. 

The resulting cross-section weighted radiation field $J_\mathrm{csw}$ impinging on the chromosphere is shown in panel d of Fig.~\ref{fig:emm_j_n_corr}. It shows a good correlation with the locations of strong emission on large scales, but not on small scales. This is because the radiation field at location $\vec{r}$ is in essence the integral of the emissivity weighted with the inverse distance squared:
\be
J_\mathrm{csw}(\vec{r}) \approx \int_0^\infty \int_V \frac{4\pi}{h \nu} \sigma_\nu \frac{\psi_\nu(\vec{r'})}{ |\vec{r}-\vec{r'}|^2 } \mathrm{e}^{ -\tau(\vec{r},\vec{r'})} \, \dd\vec{r'} \dd\nu,
\ee
with $\tau(\vec{r},\vec{r'})$ the optical thickness between  points $\vec{r}$ and $\vec{r'}$, and the integral over $\vec{r'}$ is taken over all space. The emissivity is mainly due to the corona where the extinction is negligible, and the extinction is essentially chromospheric. Therefore $J_\mathrm{csw}$ is sensitive to both strong, closely located sources as well as larger regions at larger distances. The column-integrated lower level population of the \Heline\ shows a similar large scale correlation (panel~e, the actual level population per volume is shown in blue in the animation). {In panel~f we show the resulting image at the nominal line center wavelength, which shows fine structuring up to a scale of a fraction of a Mm.}

{Note that the term $ \mathrm{e}^{ -\tau(\vec{r},\vec{r'})} $ contains significant complexity. It encodes the different opacity at different wavelengths, not only for helium but also for the main other absorber, which is the hydrogen Lyman continuum. Assuming all hydrogen and helium neutral, then the \Hecont\ makes up 40\% of the total chromospheric opacity at 50~nm and 70\% at 15~nm. Our analysis thus somewhat overestimates the effect of long-wavelength photons, because $\Lambda_\mathrm{csw} $ does not take this wavelength dependence into account}

\begin{figure}
   \centering
   \includegraphics[width=\hsize]{\figspath/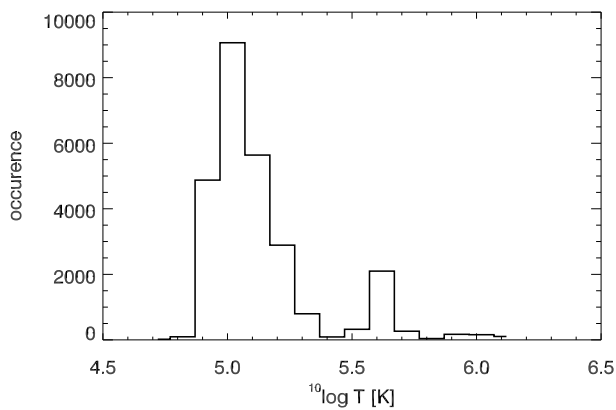}
       \caption{Histogram of the temperature of the gas for grid points in the model with $L_\mathrm{csw}> 0.01 \max(L_\mathrm{csw}))$.
      \label{fig:tdist}}
    \end{figure}

Figure~\ref{fig:emm_j_n_corr} and the animation show that the high emissivity points are located close to the chromosphere These will thus contribute most to the high $J_\mathrm{csw}$ in the nearby chromosphere.  In Fig~\ref{fig:tdist} we show the temperature distribution of the high-$L$ points. It peaks at 100 kK, indicating that the emission is coming from the transition region. Inspection of the cross-section-weighted emissivities computed with CHIANTI at $T=100$~kK and $n_\mathrm{e} =10^{16}$~m$^{-3}$ show that only 37 lines of \ion{C}{III}, \ion{C}{IV}, \ion{N}{III}, \ion{Ne}{II}, \ion{Ne}{III}, \ion{Ne}{IV}, and \ion{C}{III} produce 88\% of the ionising photons. 

In summary, we have shown that in our model the main source of {the small-scale variation of} ionizing radiation in the \Hecont\ is a thin shell of relatively dense material at ~100 kK in the transition region. Whenever a pocket of chromospheric material is sufficiently illuminated by this material it develops appreciable opacity in the \Heline. This mainly occurs in material that is almost completely surrounded by TR material, i.e., slender extensions of the chromosphere that protrude into the corona. 

\subsection{Other sources of ionizing UV radiation}

  \begin{figure*}
   \centering
   \includegraphics[width=17cm]{\figspath/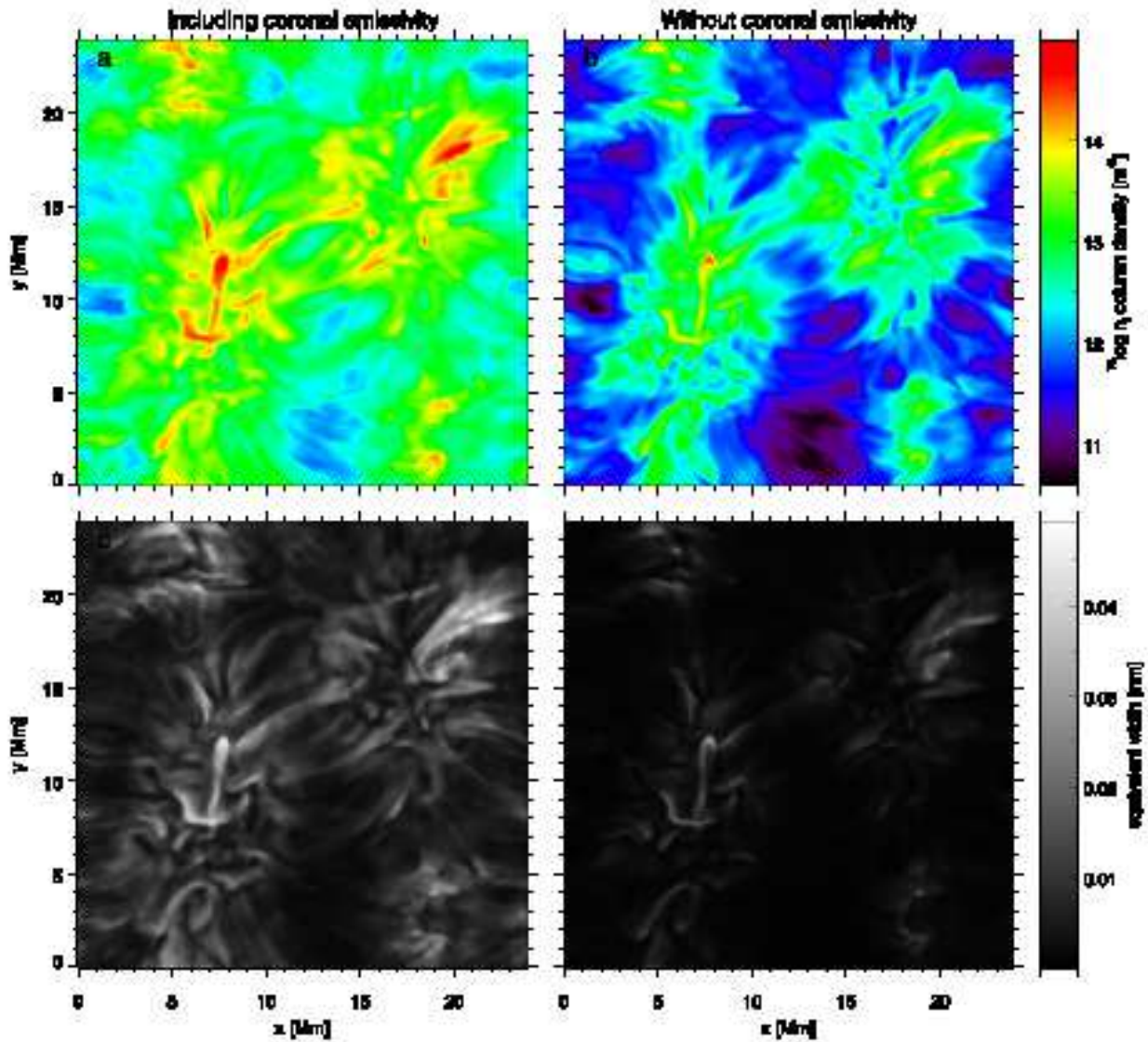}
        \caption{Demonstration of the importance of radiation from the corona and transition region for the photoionisation in the \Hecont. Panels a and b compare the column integrated lower-level population of the \Heline\ line with (a) and without (b) the inclusion of coronal and TR emission.  Panels c (with emission) and d (without emission) { compare the resulting equivalent widhts of the \Heline\ line. }        \label{fig:comp_coronarad}}
    \end{figure*}

Besides radiation from bound-bound transitions in the TR and corona, there are other sources of ionising radiation. The most important contributions are the \HeII\ ground state continuum, the \HeII~30.4 nm line and the hydrogen Lyman continuum. The first two are included self-consistently in statistical-equilibrium non-LTE because they are included in the model atom. The latter is included as a background element in the following way: the hydrogen level populations are taken from the \bifrost\ model, where they were computed including non-equilibrium ionisation effects
\citep{2016ApJ...817..125G}.
The Lyman continuum is treated as a background process in our approximation, which means we assume coherent scattering and a photon destruction coefficient computed using the Van Regemorter approximation.
 We solved the 3D non-LTE radiative transfer problem both with and without the coronal emissivities to investigate the importance of the hydrogen Lyman and  \HeII\  continua on the \Heline\ opacity.
 
Figure~\ref{fig:comp_coronarad} compares both computations. Ignoring the coronal emissivity leads to much lower total chromospheric column density, and thus optical thickness, for the \Heline, on average a factor 6.3 less (panel a and b). The resulting optical thickness is too low to leave a strong visible imprint in the intensity images at the line core. Only at a few locations is the photoionisation of \HeI\ by \HeII\ and \Hi\ radiation strong enough to be clearly visible. The best example is the upside-down T-shaped dark feature at $(x,y)=(7,8)$~Mm.

 {The frequency-integrated spatially-averaged intensity of the \HeII~30.4 nm line in our model is 13~W~m$^{-2}$~ster$^{-1}$,  which is about 27\% of the total intensity between 14.3 and 50.4~nm. This is somewhat higher than the observed values of
\citet{1999MNRAS.308..510M},
who measured 5~W~m$^{-2}$~ster$^{-1}$ for internetwork and 10~W~m$^{-2}$~ster$^{-1}$ for a strong network boundary. So while our computation of the \HeII~30.4 nm line might be deficient in the sense that we ignore non-equilibrium effects, it nevertheless produces roughly the right intensity. We thus conclude that the line is an important, but not dominant, source of ionizing photons in the \Hecont.}


\subsection{Importance of direct collisional excitation}

{We computed the $N_\mathrm{hot}$, total number of \HeI\ atoms in the lower level of the \Heline\ line in our model chomosphere, transition region and corona, for those grid points with a temperature above 20~kK, and computed $N_\mathrm{cold}$, the same quantity but  now for those points where $T<20$~kK. We find that $N_\mathrm{hot}/N_\mathrm{cold}=0.005$, and thus conclude that direct collisional excitation is an insignificant source of \Heline\ opacity in our statistical equilibrium computation.}

\subsection{3D effects on the UV illumination}

  \begin{figure*}
   \centering
   \includegraphics[width=17cm]{\figspath/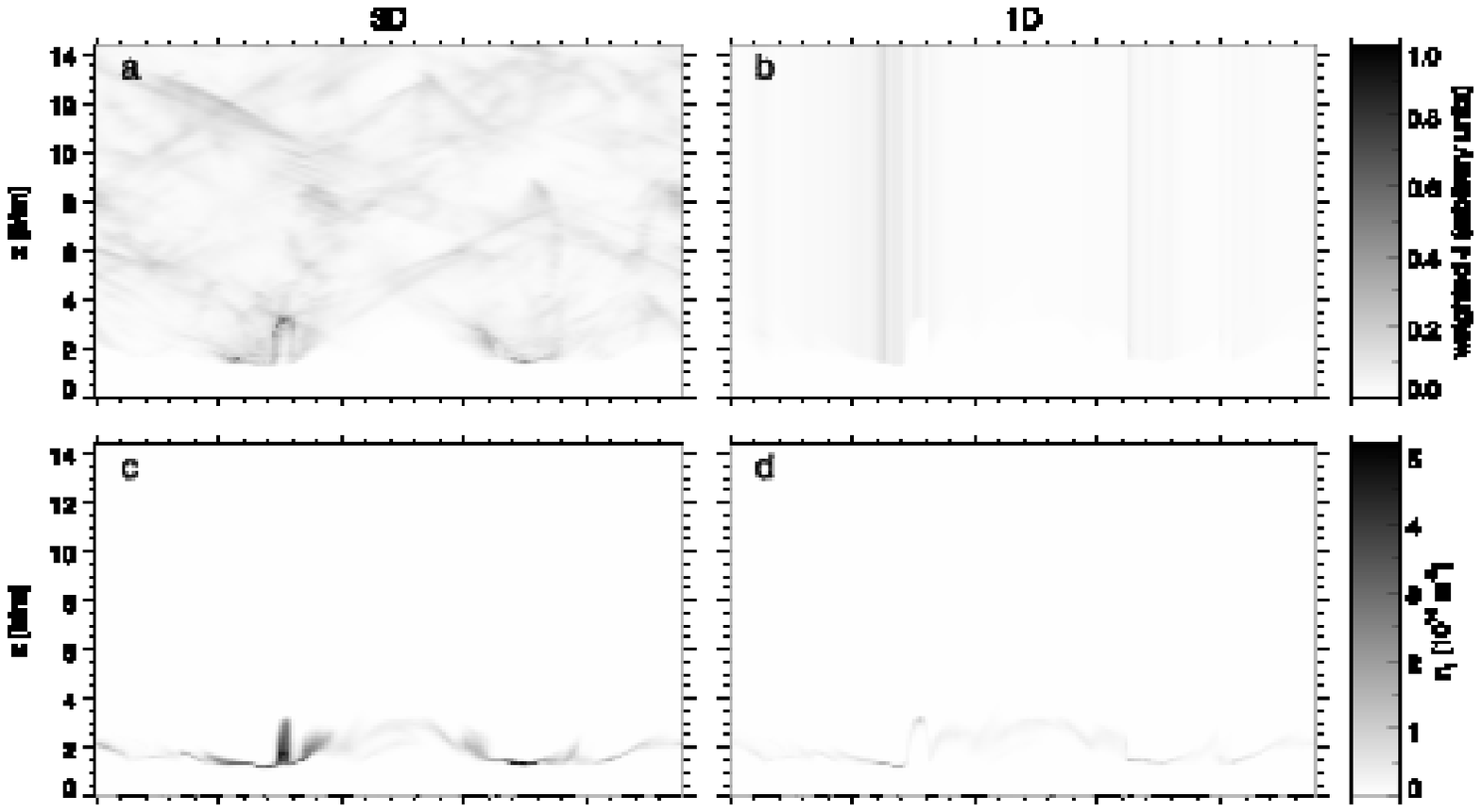}
      \includegraphics[width=17cm]{\figspath/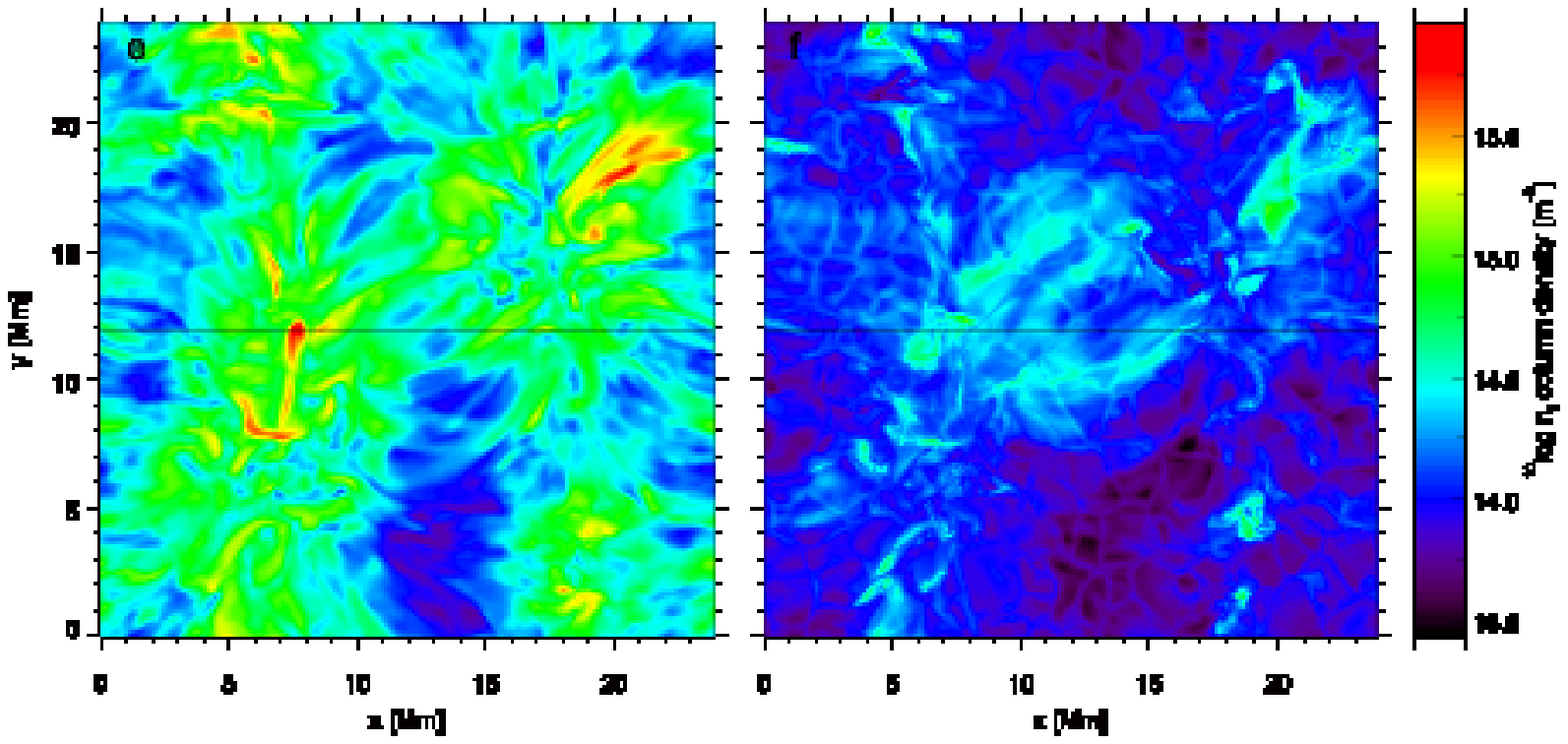}
      \caption{Comparison of the \Heline\  lower-level population assuming 3D radiative transfer (left-hand column) and 1D radiative transfer (right-hand column). Top row: cross-section weighted angle-averaged radiation field in a vertical slice at $y=11.9$~Mm. Middle row: resulting lower level population of the \Heline\ line in a vertical slice at $y=11.9$~Mm. Bottom row: column integrated lower-level population of the \Heline\ line, with the slice used in panels a---d indicated with a black line.  
      \label{fig:opac_1d_3d}}
    \end{figure*}

The coronal UV radiation spreads out in 3D. We test the importance of the horizontal spreading by comparing a computation using the full 3D radiation field (3D) and a computation where each column in the model atmosphere is treated as an independent plane-parallel atmosphere (1D). In Fig.~\ref{fig:opac_1d_3d} we show the differences between the two computations. 

Panels a and b compare the UV radiation field in a vertical slice through the model. Panel b show vertical striping in $J$, typical for 1D calculations, whereas the 3D computations has a more even distribution. Notice however that the 3D computation also shows artefacts: the diagonal striping is caused by the finite number of ray angles that we use (24 angles). A small volume with high emissivity will thus only illuminate the surrounding space along the ray directions. Because the coronal emissivity is highly structured this shows up strongly. The middle row compares the resulting \Heline\ lower level populations along the same vertical slice. In 3D the population is much higher than in 1D. This is mainly due to the effective self-shielding in 1D geometry,
because radiation can only come from one direction. In 3D extensions of chromospheric material into the corona can be illuminated from all sides. A clear example can be seen at $(x,z)=(7.5,2.0)$~Mm, where a finger of chromospheric material is surrounded by high-emissivity TR/coronal material. Similar effects were already reported by
\citet{2012A&A...539A..39C},
who investigated 3D effects on the heating of the chromosphere by UV radiation produced in the corona.

Finally, in panels e and f we compare the spatial distribution of the total column density of the lower-level population of the \Heline\ line. In 3D it is on average 4.2 times higher than in 1D, and its maximum is 6.8 times higher.

\subsection{3D effects on the source function} \label{sec:3DS}

  \begin{figure}
   \centering
   \includegraphics[width=\hsize]{\figspath/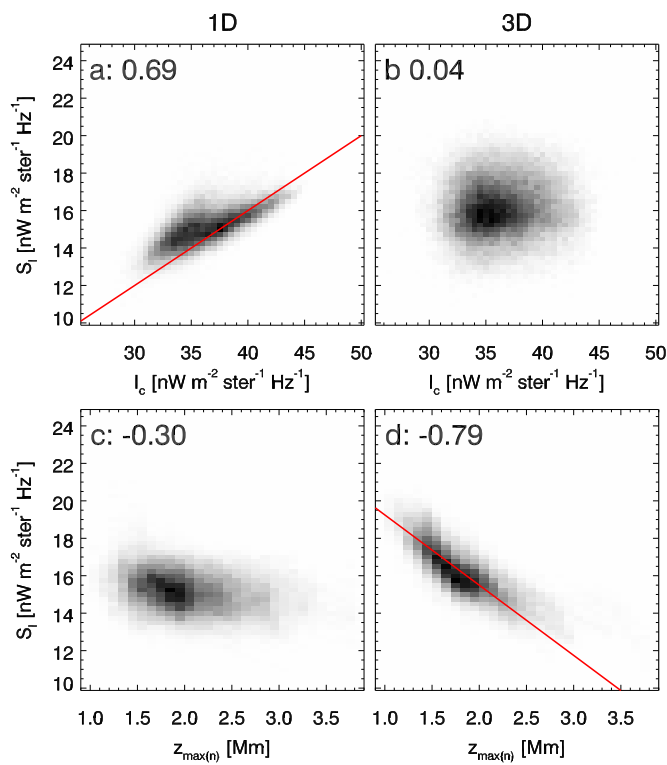}
       \caption{JPDFs of the \Heline\ source function and the continuum intensity and the height where the line opacity is largest.  The red line in panel a is $S_\mathrm{l} = 0.4 I_\mathrm{c}$. The red line in panel d is $S_\mathrm{l} = 23- 3.75 z_{\max{(n)}}$
      \label{fig:Scorr}}
    \end{figure}

At the height in the chromosphere where the \Heline\ line has appreciable opacity, the mass density is so low that the damping wings of the Voigt profile are negligible, and the extinction coefficient is essentially Gaussian. The photon destruction probability $\epsilon$ in the \Heline\ is of the order $10^{-4} - 10^{-2}$. The line is thus scattering dominated ($S_\nu \simeq \bar{J}_{\nu0}$) and the thermalisation depth is located at a line core optical depth $\tau \approxeq 1/\epsilon = 10^2$ -- $10^4$. This is at least eleven times the maximum optical thickness of the line-forming region in our model, which is  8.9. So the source function thermalises in the photosphere, even when the line-forming region in the chromosphere is optically thick. We thus expect strong 3D effects on the source function. 

We investigated this effect by comparing the full 3D computation with the 1D column-by-column computation. The top row of Figure~\ref{fig:Scorr} shows joint probability density functions  (JPDFs) of the line source function $S_\mathrm{l}$ and the underlying continuum intensity $I_\mathrm{c}$ computed from each $(x,y)$-pixel in our model, for both the 3D and 1D computation.

Panel a shows a strong correlation between $S_\mathrm{l}$ and $I_\mathrm{c}$ in the 1D computation. The red line shows that roughly speaking, $S_\mathrm{l} = 0.4 I_\mathrm{c}$. This is consistent with
\citet{1994isp..book...35A}
who obtained the same relation for 1D semi-empirical models illuminated with a prescribed UV radiation field. In 3D this correlation is gone (panel b). The reason is that the source function at a given location in the chromosphere is no longer correlated to the continuum intensity from the photosphere directly below it, but by a spatial average of the continuum intensity emitted from an extended region encompassing multiple granules.

The bottom row of Figure~\ref{fig:Scorr} shows the JPDFs for $S_\mathrm{l}$  and the height where the lower-level population is maximum in the chromosphere $z_{\max{(n)}}$. In 1D there is just a rather weak correlation between the two quantities, which only shows up in the correlation coefficient. In 3D it is strong, and is clearly visible in the JPDF as well.

The behaviour of the \Heline\ source function is similar to the source function in the \Ha\ line
\citep{2012ApJ...749..136L}.
Both lines have source functions that thermalise in the photosphere, both lines are strongly scattering and both lines have negligible opacity in the upper photosphere and lower chromosphere.

\subsection{Relative importance of 3D in the UV illumination and source function}

  \begin{figure}
   \centering
   \includegraphics[width=\hsize]{\figspath/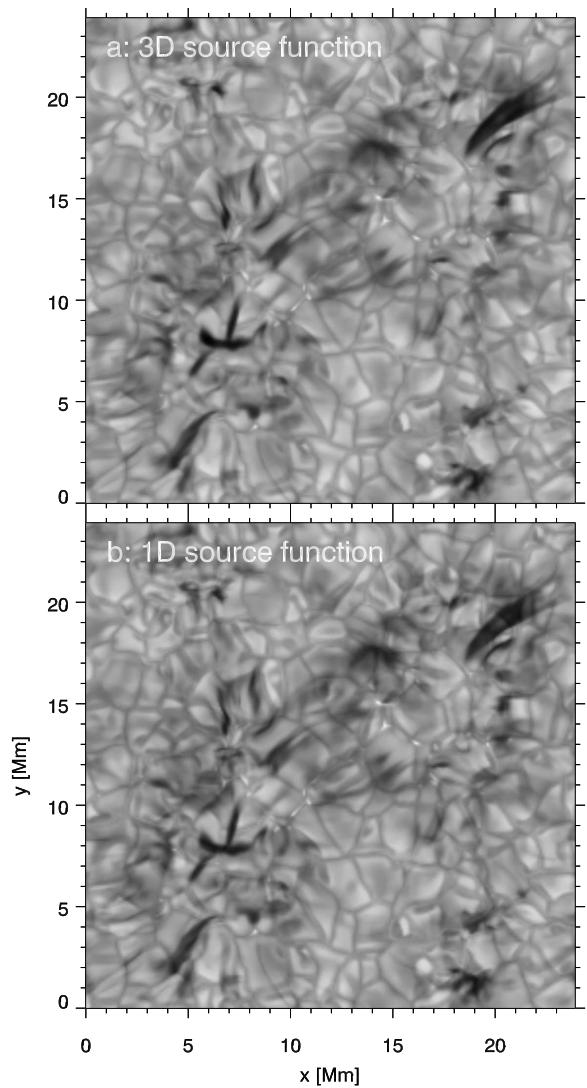}
       \caption{Comparison of the synthetic \Heline\ line intensity computed with the correct 3D source function (upper panel) and an approximate 1D source function in the chromosphere (lower panel).
             \label{fig:3DStest}}
    \end{figure}

In order to assess the relative importance of 3D effects in the UV illumination and the source function we did a test calculation where we computed the emergent intensity using the opacity from the 3D computation, but used a line source function of the form
\be
S_\mathrm{l}(x,y,z) = 0.4 I_\mathrm{c}(x,y),
\ee
which  was shown in Section~\ref{sec:3DS} to be a decent approximation of the source function as obtained from 1D plane-parallel computations. The resulting line-center intensity image, together with the 3D computation, is shown in Fig~\ref{fig:3DStest}. Both images are very similar in absorption structure, so we conclude that 3D effects in the UV illumination are the most important for the \Heline\ formation, while 3D effects on the line source function are minor.

\subsection{Variation in \Heline\ {opacity} and the chromospheric electron density}

  \begin{figure*}
   \centering
  \includegraphics[width=17cm]{\figspath/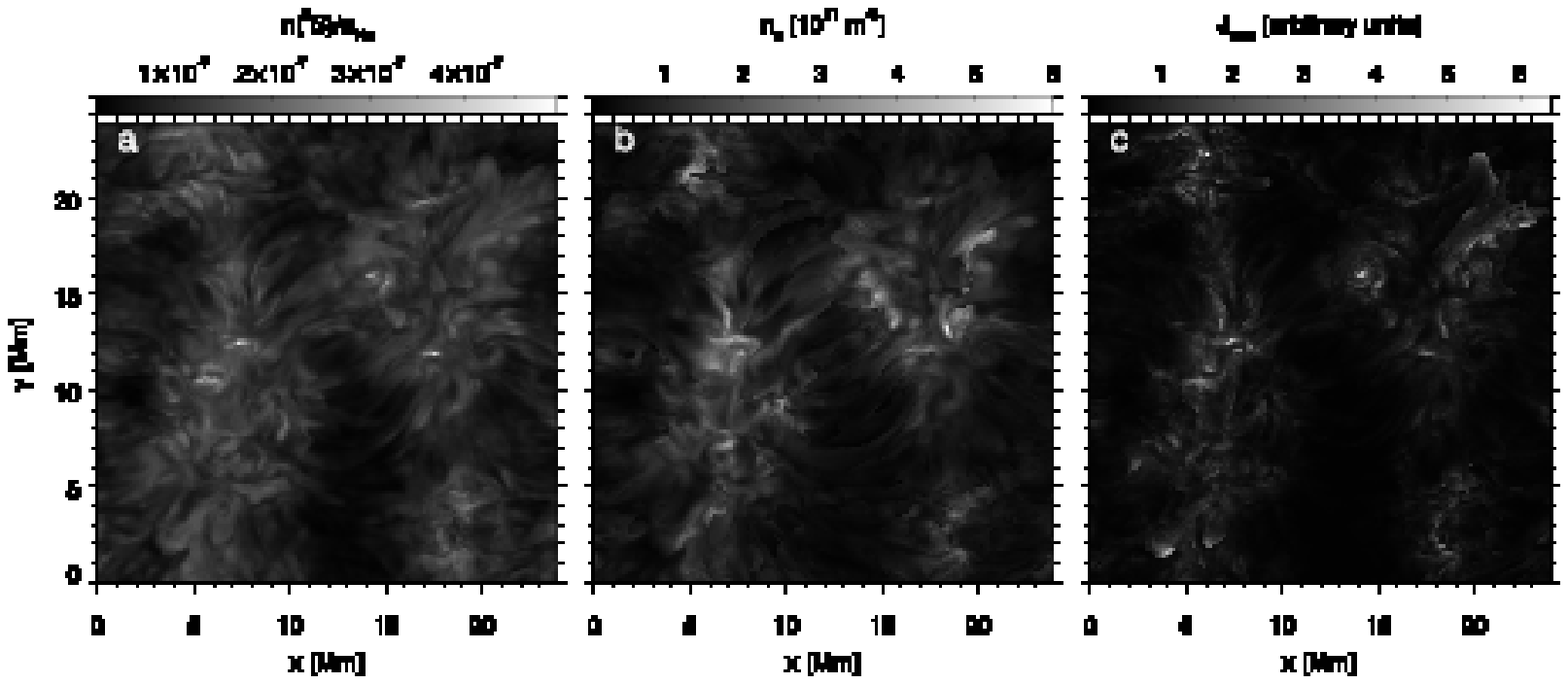}
    \includegraphics[width=17cm]{\figspath/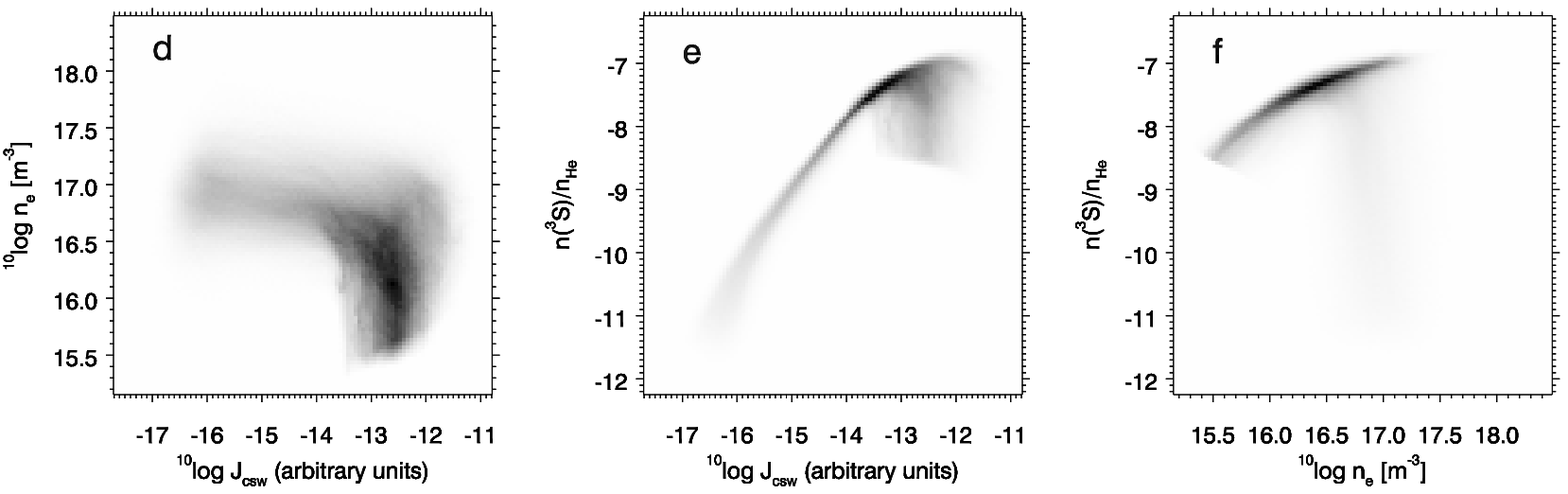}
  \caption{Relation between \Heline\ opacity, electron density and the coronal irradiation. Panel a: relative population of the \Heline\ lower level, for each pixel at the height in the chromosphere where the relative population has a maximum. Panel b: electron density at the same location as in a. Panel c: $J_\mathrm{csw}$ at the same location as in a and b. Panel d: JPDF of $J_\mathrm{csw}$  and the electron density in the chromosphere. Panel d: JPDF of $J_\mathrm{csw}$ and the relative population of the \Heline. Panel f: JPDF of the electron density and the relative population of the \Heline.
  \label{fig:edens_corr}}
    \end{figure*}

In the previous subsections we explored the origin and effect of the ionising radiation. In this subsection we discuss the combined influence of the electron density and the radiation. Equation~\ref{eq:relpop}, which was derived by
\citet{1976SoPh...49..315L},
predicts a correlation between the relative population \relpop\  of the \Heline, and those quantities. We illustrate this in Fig.~\ref{fig:edens_corr}. For each $(x,y)$ location in the model we looked up the location in the chromosphere where \relpop\ has a maximum. For those locations we extracted \relpop, \ne, and $J_\mathrm{csw}$ and display the resulting maps in panels a--c.  The map of  \relpop\ is finely structured on small scales, but appears relatively smooth on larger scales. The electron density map shows a substantial correspondence with the \relpop\ map. The map of $J_\mathrm{csw}$ in panel c is, in contrast, much more intermittent.

In panels d--f we put the correlation on a more precise footing. Panel d shows the joint probability density function of $J_\mathrm{csw}$ and \ne. It shows a peculiar $\neg$-shape. Panel e shows the JPDF of $J_\mathrm{csw}$ and \relpop. It consists of two distinct components, a tight positive correlation that roughly follows the prediction of Eq.~\ref{eq:relpop} for fixed electron density and a beard-shaped cluster of points of $-13 < \log_{10} J_\mathrm{csw} < -12$. In panel d it can be seen that the latter points correspond to a sudden large spread in values of the electron density, and thus a much weaker correlation between $J_\mathrm{csw}$ and \relpop. 
A similar effect appears in the JPDF of \ne\ and \relpop: there is a component with a tight correlation and for $ 16.5 <\log_{10} \ne< 17$ there is a beard-like extension downward, caused by the large spread of values in $J_\mathrm{csw}$ at those electron densities, as shown in Panel d.

We thus conclude that the approximate formula made by 
\citet{1976SoPh...49..315L}
is validated by our 3D non-LTE radiative transfer computations.

\section{Correlation with \ion{Si}{IV} 1393 and 1402 and AIA 304, 171,193 and 335 emission} \label{sec:test}

  \begin{figure}
   \centering
   \includegraphics[width=\hsize]{\figspath/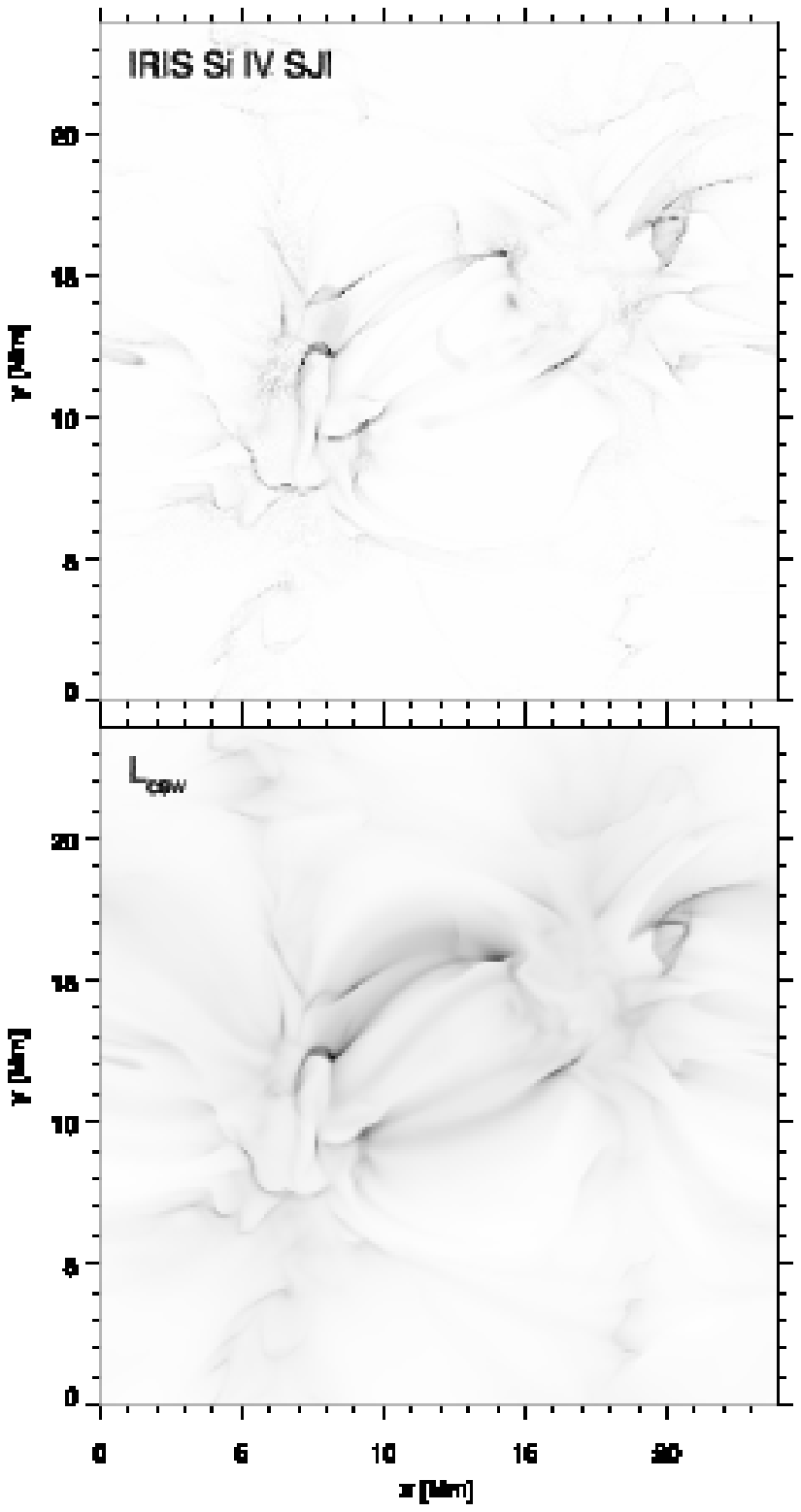}
   \includegraphics[width=\hsize]{\figspath/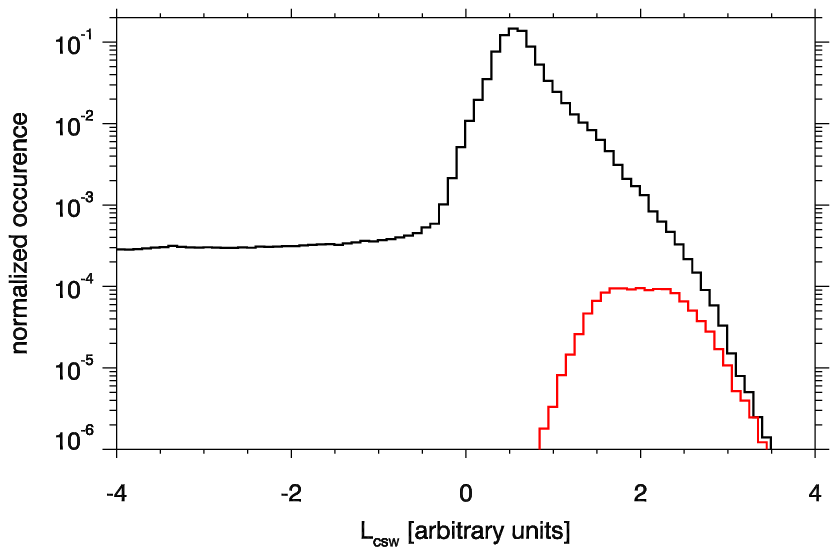}
       \caption{Comparison of the synthetic IRIS 140~nm slit jaw emission (upper panel) and an image computed from $L_\mathrm{csw}$, the frequency integrated and cross-section weighted photon losses (middle panel). The bottom panel shows the histogram of  $L_\mathrm{csw}$ for all grid points in the simulation (black) and only for the 0.1\% of the grid points that have the highest \SiIV\ emissivity (red).
                 \label{fig:SiIV_Lcsw}}
    \end{figure}

  \begin{figure*}
   \centering
  \includegraphics[width=17cm]{\figspath/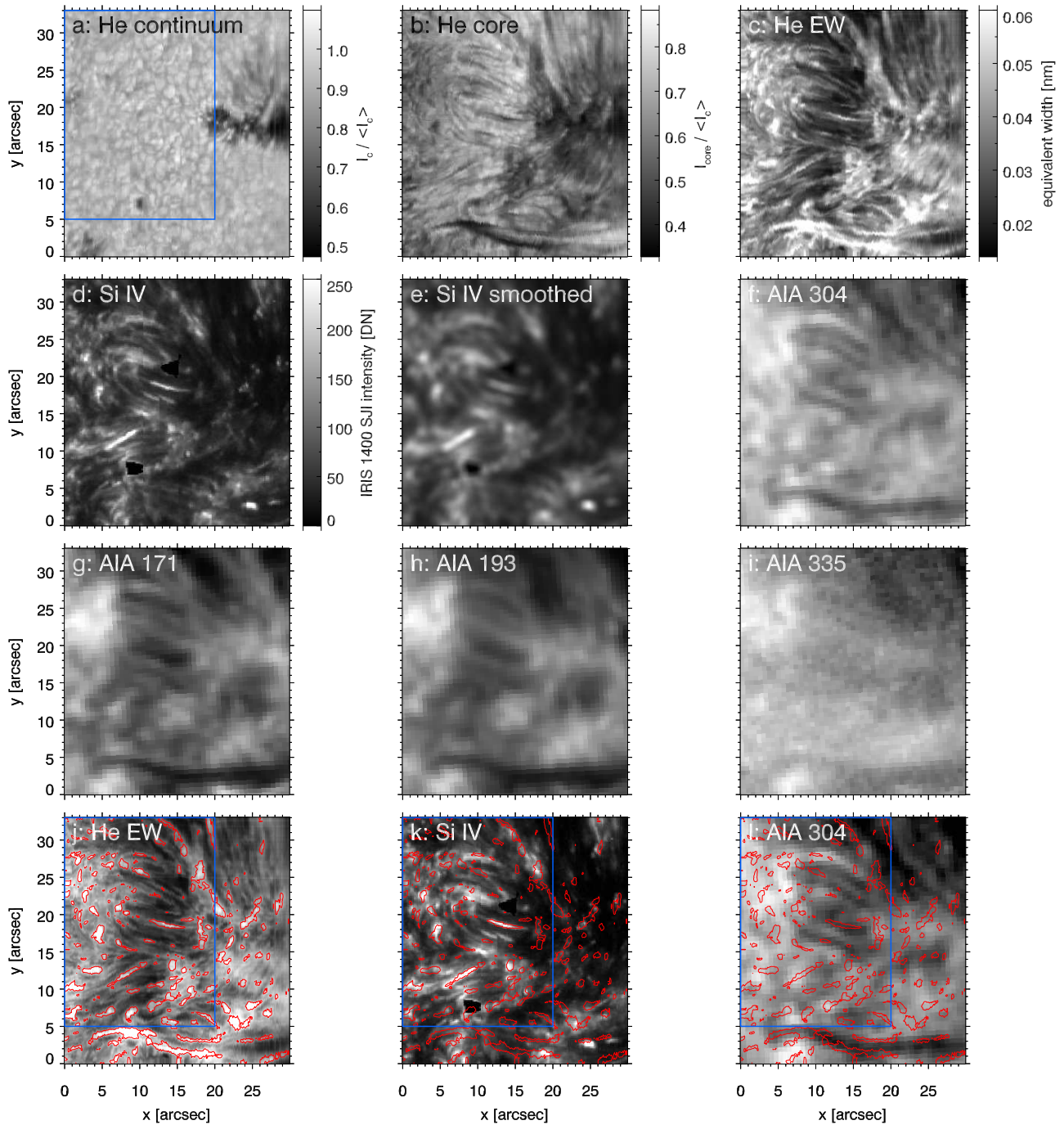}
  \caption{Co-spatial and co-temporal observations of active region NOAA 12393 with the SST, IRIS and SDO/AIA. a) TRIPPEL raster scan image in the continuum at 1081.971~nm. b)  TRIPPEL raster scan image of the profile minimum of \Heline. c) TRIPPEL raster scan image of the equivalent width of \Heline. d) artificial raster scan constructed from the IRIS 140 nm SJI images which are dominated by \SiIV\ emission. {e) same as d, but now smoothed to AIA resolution. f)--i) artificial raster scans for the AIA band given in the panels. j)--k): same as c, d, and f, but now with contours of local maxima in the EW overplotted in red. The blue box outlines the area discussed in the text.}\label{fig:Trippel-IRIS}}
    \end{figure*}

  \begin{figure}
   \centering
  \includegraphics[width=\hsize]{\figspath/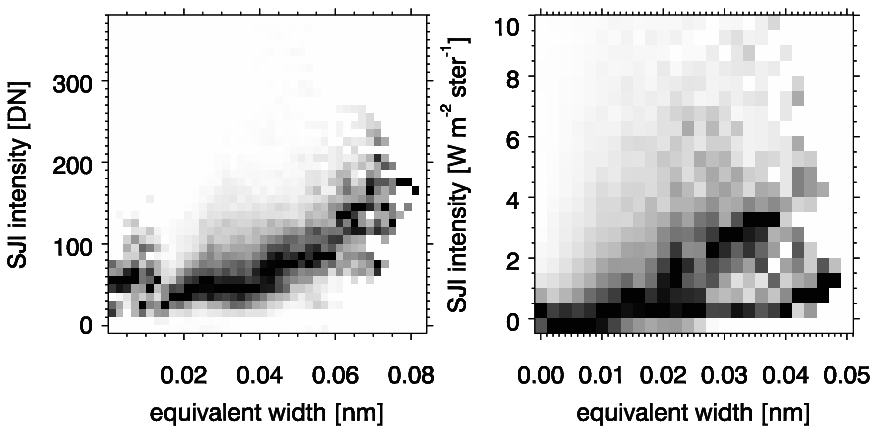}
  \caption{Joint probability density functions of the equivalent width of \Heline\ and  \SiIV 140~nm emission. Left-hand panel: observations; right-hand panel: simulations. \label{fig:EW_Si_corr}}
    \end{figure}

In Sec.~\ref{sec:coronal_sources} we showed that the source of the most intense \Hecont\ radiation impinging on the chromosphere in our simulation comes from gas with a temperature of about 100~kK. This temperature is  close to the coronal-equilibrium peak formation temperature of the \SiIV\ lines at 139.38~nm and 140.28~nm, which is at $\log T =4.8$~K. These lines dominate the emission in the 140 nm slit-jaw imager of IRIS.

In Fig.~\ref{fig:SiIV_Lcsw} we compare a synthetic \SiIV\ SJI image computed from our simulation snapshot and a map of the column-integrated $L_\mathrm{csw}$, i.e, $\int L_\mathrm{csw}\, \dd z$. The brightest  \SiIV\ SJI emission corresponds to the brightest $L_\mathrm{csw}$, but the latter is more extended and has a diffuse component, owing to the large variety of lines that contribute to it. 

The bottom panel of Fig~\ref{fig:SiIV_Lcsw} shows a histogram of $L_\mathrm{csw}$ for all grid points in the simulation in black. In red it shows a histogram of  $L_\mathrm{csw}$  for only those grid points that belong to the 0.1\% grid points that have the highest \SiIV\ SJI emmisivity. This confirms quantitatively that locations with very high  \SiIV\ emission indeed correspond to locations with the highest $L_\mathrm{csw}$ .

Based on the results shown in Fig.~\ref{fig:SiIV_Lcsw} and the rough correlation between $L_\mathrm{csw}$ and the lower-level population of \Heline\ (see Fig.~\ref{fig:emm_j_n_corr}), {we thus expect a considerable correlation between \SiIV\ emission and  \Heline\ strength on scales of $\sim 1$~arcsecond. We also show that the bulk of the ionizing photons are coming from hotter gas, emitted in a large volume in the overlying corona. We thus also expect a correlation between \Heline\ strength and coronal emission at larger scales. The latter correlation has of course been known for a long time
\citep[e.g.,][]{1983SoPh...87...47K}.
}

We therefore compared a single raster scan of the \Heline\ line taken with the TRIPPEL spectrograph at the SST with co-temporal and co-spatial artificial raster scans constructed from \SiIV\ slit-jaw images and SDO/AIA images (see Sec.~\ref{sec:observations}). Observationally we cannot directly determine atomic level populations, so we took the equivalent width (EW) as a measure instead. 

Panel~a in Fig.~\ref{fig:Trippel-IRIS} shows the raster scan in a continuum wavelength close to the \Heline\ line. There is a small sunspot with an irregular penumbra on the right side of the field-of-view, with granulation and some small pores filling the rest of the panel. Inspection of the IRIS 283.2 \MgII\, k wing slit-jaw image (not shown) indicates that the observed area shows a considerable amount of magnetic bright points. Panel~b shows an image of the \Heline\ core. In each pixel the intensity of the profile minimum is taken to compensate for Doppler shifts. The superpenumbra is clearly seen on the right side. The middle part of the image ($5<x<20 \arcsec$) is filled with fibrils with granulation visible in between. The left side of the image shows a more irregular absorption structure. The strong elongated absorption feature at the bottom of the image ($y<5\arcsec$) is an active region filament. Panel~c shows the EW of the \Heline\ line. The filament has the highest EW, while the areas showing granulation in Panel~b have the lowest. Panel~d shows the \SiIV\ artificial raster scan. The sunspot is very dark, the filament is not visible, the rest of the image shows filamentary structures with medium brightness and more irregularly shaped patches of high brightness. 

The simulation results presented in this paper are not representative for sunspots or filaments. The correlation that we expect on the basis of our simulation and radiative transfer modeling is most likely valid for quiet sun and network areas. We therefore focus on the most quiet area in the observation, roughly speaking where $x<20\arcsec$ and $y>5\arcsec$ {(indicated by the blue box in Fig.~\ref{fig:Trippel-IRIS}).}

Comparison of the brightest areas in panel d with the profile-minimum intensity and EW of  \Heline\ in panel~\ref{fig:Trippel-IRIS}b and~\ref{fig:Trippel-IRIS}c shows many places where \SiIV\ brightness coincides with a large \Heline\ EW, for example at $(x,y)=(14\arcsec,19\arcsec)$, $(x,y)=(7\arcsec,12\arcsec)$and $(x,y)=(3\arcsec,23\arcsec)$. Careful inspection shows many more locations of good correlation. There are also locations where the correlation is not present, for example at $(x,y)=(10\arcsec,23\arcsec)$ and $(x,y)=(11\arcsec,28\arcsec)$.

{Panels e shows the \SiIV\ emission, smoothed to the same resolution as the AIA data to facilitate a more fair comparison to the emission in the various AIA channels shown in panels f--i, from the coolest 304 channel to the hot 335 channel.}

{Comparing the AIA 304 channel (panel d, which is most sensitive the \HeII\ 30.4~nm emission) to \SiIV\ in panel c, we see that some structures can be recognized in both images, and there is a correlation on large scales. The \SiIV\ image shows considerable more contrast and shows many more small patches of large brightness. The difference in appearance is not only caused by differences in spatial resolution between the IRS and SDO data.}
{The large-scale patches of high EW in panel c correlate roughly with the high-brightness patches in 171 and 193, and to a lesser extent to the 304 and 335 images.}

{The bottom row (panels j--l) show the EW, \SiIV, and 304 image again, but now overlaid with contours enclosing areas of local maxima in EW. These contours where created by thresholding a high-pass-filtered EW image, designed to pick up local maxima in EW, and serve to guide the eye. Inspection of panels k and l shows that the contours of local maxima in EW match the \SiIV\ image much better than the 304 image.}

Figure~\ref{fig:EW_Si_corr} compares the joint probability density function of the \Heline\ equivalent width and the \SiIV\ brightness. For the observations we computed it from the area where $x<20\arcsec$ and $y<5\arcsec$ in Fig.~\ref{fig:Trippel-IRIS}. For the simulation we degraded the synthetic \SiIV\ image to the spatial resolution of the observations. The observed correlation clearly shows the correlation between high equivalent width and \SiIV\ brightness. For the simulations this correlation is not as strong.  The correlation appears as the upper arm of the bifurcated distribution for equivalent widths larger than 0.22~nm.

\section{Discussion and Conclusions} \label{sec:conclusions}

In this paper we investigated why the absorption features in images taken in the \Heline\ line in quiet Sun and network regions show intricate fine structuring {to sub-arcsecond scale using a 3D radiation-MHD simulation of the solar atmosphere, combined with 3D non-LTE statistical equilibrium radiative transfer modeling.}

We confirm the mechanism for populating the lower level of the line through photoionisation from the singlet ground state of  \HeI\ to \HeII, followed by photorecombination into the \HeI\ triplet system as was originally put forward by 
\citet{1939ApJ....89..673G}. 
{Direct collisional excitation at temperatures above 20~kK is a negligible source of lower level population in our model.}
We also confirm the conclusion by 
\citet{1976SoPh...49..315L} 
that the relative opacity of  \Heline\ is electron density dependent: for a given mass density and coronal illumination the extinction coefficient is an increasing function of the electron density under typical chromospheric conditions.

Our simulations are not representative for strongly magnetically active areas and phenomena, such as emerging flux regions, flares and plage. These are typically associated with elevated chromospheric densities and temperatures and the presence of non-thermal electrons, where population of the triplet system can also occur through direct collisional excitation or collisional ionisation/recombination pathways
\citep[e.g.,][]{2005A&A...432..699D}.
Our conclusions are therefore not necessarily valid under such circumstances.
 
We found that the sources of the ionising radiation can be roughly split in two components. The majority of the ionising photons are produced by gas in the 0.5-2~MK temperature range. The  source regions are extended as these are the temperatures of typical coronal gas, and their emissivity is relatively low. 
The second component consists of transition region gas between 80~kK and 200~kK. It produces far fewer ionising photons than the high-temperature component. However, regions of high emissivity of the second component are concentrated in space and are located close to the upper chromosphere. This second, highly structured component is the main cause of variation in the UV flux impinging on the chromosphere. The actual radiation field $J_\mathrm{csw}$ inside the chromosphere also depends on the shape of the transition region. Chromospheric material that is surrounded by a dense transition region has the highest $J_\mathrm{csw}$ and consequently the highest \Hecont\ photoionisation rate.

{The \Heline\ line strength depends on the integral of the opacity along the line of sight. Therefore dense, vertically-oriented chromospheric structures that are surrounded by hot, dense, transition region and coronal material will typically produce the strongest line.}

We tested our predicted correlation between the presence of high-mass-density $\approx100$~kK gas and  \Heline\ absorption by comparing emission in the \SiIV\ 139.38~nm and 140.28~nm lines that form around 80~kK with \Heline\ absorption {and found that small-scale patches of high \Heline\ EW} indeed often correlate with high \SiIV\ emission. The correlation is notably absent around the active region filament in our observations, which might indicate a thin and/or low-mass-density transition region around the filament. {The correlation between small-scale structure in the EW and AIA~304 is significantly worse. The large-scale variation in EW correlates well with the emission in coronal lines as sampled by the AIA 171 and 193 as well as the AIA 304 emission.}


Our work is an improvement over earlier modelling efforts
\citep[e.g.,][]{1975ApJ...199L..63Z,1988SvA....32..542P,1994isp..book...35A,2008ApJ...677..742C}.
We used a 3D radiation-MHD simulation of the solar atmosphere, as opposed to the 1D models used before. We self-consistently include the emission of ionising radiation from the corona: it is produced by the corona present in the radiation-MHD simulation. Earlier work used a prescribed coronal radiation field.

Despite these improvements there are still a number of limitations to our approach:  The simulation that we use produces too little chromospheric dynamics and produces to little chromospheric emission compared to observations
\citep[see for example][]{2016A&A...585A...4C, 2013ApJ...772...90L}.
Increasing the numerical resolution and adding heating effects from ion-neutral interactions
\citep{2012ApJ...753..161M,2012ApJ...747...87K}
might alleviate some of the discrepancy in future simulations. We also base our current analysis on a single snapshot, with a single magnetic field configuration. A future improvement would be to repeat the analysis for a time series of snapshots and for simulation runs with different magnetic field geometries.

The biggest discrepancy in our current analysis is that we do not include the effect of non-equilibrium ionisation in our 3D non-LTE radiative transfer calculation for helium, nor do we include non-equilibrium ionisation effects on the production of ionising photons. 

Non-equilibrium ionisation in the transition region and corona typically has the effect that ions in a particular ionisation stage are present over a larger temperature range than what one would expect from coronal equilibrium calculations
\citep{2013AJ....145...72O}.
The effect on the resulting emissivities and intensities are rather mild for the lines of \SiIV, \ion{C}{IV}, and \ion{O}{IV} that were investigated by
\citet{2015ApJ...802....5O}.
No non-equilibrium intensity calculations for the lines dominating the \HeI\ photoionisation mentioned in {at the end of Sec.~\ref{sec:coronal_sources}} have been done, but the results for the non-equilibrium ionisation balance of oxygen from
\citet{2013ApJ...767...43O}
indicate that \ion{O}{III} lines should experience similar mild effects as the \ion{O}{IV} lines. {Non-equilibrium effects also play a role for the \HeII\ 30.4~nm line 
\citep[e.g.][]{2014ApJ...784...30G}.
However, we showed that the line produces roughly the right amount of photons, so that neglecting this effect has little effect on the formation of the \Heline\ line.}

Comparisons of one-dimensional calculations assuming SE and NE for helium using the \radyn\ code
\citep[e.g.,][]{1992ApJ...397L..59C}
indicate that large differences in the \Heline\ profiles can occur
\citep{2014ApJ...784...30G}.
While we expect that inclusion of non-equilibrium effects will have large quantitative effects on \Heline\ formation, we do not think it will have a qualitative effect on the spatial structure: also in the non-equilibrium case the photoionisation-recombination mechanism {is the dominant mechanism} populating the triplet states
\citep{2014ApJ...784...30G},
The correlation between \Heline\ absorption and strong emission in the transition region and high chromospheric electron density will thus remain valid, {even though the contribution of direct collisional excitation due to the presence of neutral helium at higher temperatures might play a bigger role.}
Because we expect quantitative differences when non-equilibrium radiative transfer for helium is taken into account we have refrained from detailed analysis of the synthetic \Heline\ profiles in this work. 

\begin{acknowledgements}
 This research was supported by the Research Council of Norway through
 the grant ``Solar Atmospheric Modelling'' and through grants of
 computing time from the Programme for Supercomputing. The research leading to these results has received funding from the European Research Council under the European Union's Seventh Framework Programme (FP7/2007-2013) / ERC grant agreement no 291058. Some 
 computations were performed on resources provided by the Swedish 
 National Infrastructure for Computing (SNIC)  at the National Supercomputer 
 Centre at Link\"oping University. The Swedish 1-m Solar Telescope is operated on the island of La Palma
by the Institute for Solar Physics of the Royal Swedish Academy of
Sciences in the Spanish Observatorio del Roque de los Muchachos of the
Instituto de Astrof{\'\i}sica de Canarias. IRIS is a NASA small explorer mission developed and operated by LMSAL with mission operations executed at NASA Ames Research center and major contributions to downlink communications funded by ESA and the Norwegian Space Centre.
 
\end{acknowledgements}


\bibliographystyle{aa} 
\bibliography{%
auer,%
andretta,%
avrett,%
bradshaw,%
carlson,%
carlsson,%
delacruz,%
depontieu,%
dere,%
fontenla,%
goldberg,%
gustafsson,%
hansteen,%
harvey-jack,%
hubeny,%
kiselman,%
khomenko,%
joselyn,%
judge,%
landi,%
leenaarts,%
martinez-sykora,%
mihalas,%
olluri,%
osterbrock,%
pereira,%
rybicki,%
schad,%
scharmer,%
temp,%
tobiska,%
pozhalova,%
uitenbroek,%
zirin%
}

\begin{thebibliography}{59}
\expandafter\ifx\csname natexlab\endcsname\relax\def\natexlab#1{#1}\fi

\bibitem[{{Andretta} \& {Jones}(1997)}]{1997ApJ...489..375A}
{Andretta}, V. \& {Jones}, H.~P. 1997, \apj, 489, 375

\bibitem[{{Auer}(2003)}]{2003ASPC..288....3A}
{Auer}, L. 2003, in ASP Conf. Ser. 288: Stellar Atmosphere Modeling, 3

\bibitem[{{Avrett} {et~al.}(1994){Avrett}, {Fontenla}, \&
  {Loeser}}]{1994isp..book...35A}
{Avrett}, E.~H., {Fontenla}, J.~M., \& {Loeser}, R. 1994, {Formation of the
  solar 10830 A line} (Infrared Solar Physics, IAU Symp.~No.~154,
  eds.~D.M.~Rabin, J.T.~Jefferies, and C.~Lindsey, Kluwer, Dordrecht,
  pp.~35-47), 35--47

\bibitem[{{Bradshaw} \& {Raymond}(2013)}]{2013SSRv..178..271B}
{Bradshaw}, S.~J. \& {Raymond}, J. 2013, \ssr, 178, 271

\bibitem[{Carlson(1963)}]{carlson1963}
Carlson, B.~G. 1963, Methods of Computational Physics, 1

\bibitem[{{Carlsson} {et~al.}(2016){Carlsson}, {Hansteen}, {Gudiksen},
  {Leenaarts}, \& {De Pontieu}}]{2016A&A...585A...4C}
{Carlsson}, M., {Hansteen}, V.~H., {Gudiksen}, B.~V., {Leenaarts}, J., \& {De
  Pontieu}, B. 2016, \aap, 585, A4

\bibitem[{{Carlsson} \& {Leenaarts}(2012)}]{2012A&A...539A..39C}
{Carlsson}, M. \& {Leenaarts}, J. 2012, \aap, 539, A39

\bibitem[{{Carlsson} \& {Stein}(1992)}]{1992ApJ...397L..59C}
{Carlsson}, M. \& {Stein}, R.~F. 1992, \apjl, 397, L59

\bibitem[{{Carlsson} \& {Stein}(2002)}]{2002ApJ...572..626C}
{Carlsson}, M. \& {Stein}, R.~F. 2002, \apj, 572, 626

\bibitem[{{Centeno} {et~al.}(2008){Centeno}, {Trujillo Bueno}, {Uitenbroek}, \&
  {Collados}}]{2008ApJ...677..742C}
{Centeno}, R., {Trujillo Bueno}, J., {Uitenbroek}, H., \& {Collados}, M. 2008,
  \apj, 677, 742

\bibitem[{{Cirtain} {et~al.}(2013){Cirtain}, {Golub}, {Winebarger}, {De
  Pontieu}, {Kobayashi}, {Moore}, {Walsh}, {Korreck}, {Weber}, {McCauley},
  {Title}, {Kuzin}, \& {Deforest}}]{2013Natur.493..501C}
{Cirtain}, J.~W., {Golub}, L., {Winebarger}, A.~R., {et~al.} 2013, \nat, 493,
  501

\bibitem[{{de la Cruz Rodr{\'{\i}}guez} {et~al.}(2015){de la Cruz
  Rodr{\'{\i}}guez}, {L{\"o}fdahl}, {S{\"u}tterlin}, {Hillberg}, \& {Rouppe van
  der Voort}}]{2015A&A...573A..40D}
{de la Cruz Rodr{\'{\i}}guez}, J., {L{\"o}fdahl}, M.~G., {S{\"u}tterlin}, P.,
  {Hillberg}, T., \& {Rouppe van der Voort}, L. 2015, \aap, 573, A40

\bibitem[{{De Pontieu} {et~al.}(2014){De Pontieu}, {Title}, {Lemen}, {Kushner},
  {Akin}, {Allard}, {Berger}, {Boerner}, {Cheung}, {Chou}, {Drake}, {Duncan},
  {Freeland}, {Heyman}, {Hoffman}, {Hurlburt}, {Lindgren}, {Mathur}, {Rehse},
  {Sabolish}, {Seguin}, {Schrijver}, {Tarbell}, {W{\"u}lser}, {Wolfson},
  {Yanari}, {Mudge}, {Nguyen-Phuc}, {Timmons}, {van Bezooijen}, {Weingrod},
  {Brookner}, {Butcher}, {Dougherty}, {Eder}, {Knagenhjelm}, {Larsen},
  {Mansir}, {Phan}, {Boyle}, {Cheimets}, {DeLuca}, {Golub}, {Gates}, {Hertz},
  {McKillop}, {Park}, {Perry}, {Podgorski}, {Reeves}, {Saar}, {Testa}, {Tian},
  {Weber}, {Dunn}, {Eccles}, {Jaeggli}, {Kankelborg}, {Mashburn}, {Pust},
  {Springer}, {Carvalho}, {Kleint}, {Marmie}, {Mazmanian}, {Pereira}, {Sawyer},
  {Strong}, {Worden}, {Carlsson}, {Hansteen}, {Leenaarts}, {Wiesmann},
  {Aloise}, {Chu}, {Bush}, {Scherrer}, {Brekke}, {Martinez-Sykora}, {Lites},
  {McIntosh}, {Uitenbroek}, {Okamoto}, {Gummin}, {Auker}, {Jerram}, {Pool}, \&
  {Waltham}}]{2014SoPh..289.2733D}
{De Pontieu}, B., {Title}, A.~M., {Lemen}, J.~R., {et~al.} 2014, \solphys, 289,
  2733

\bibitem[{{Dere} {et~al.}(1997){Dere}, {Landi}, {Mason}, {Monsignori Fossi}, \&
  {Young}}]{1997A&AS..125..149D}
{Dere}, K.~P., {Landi}, E., {Mason}, H.~E., {Monsignori Fossi}, B.~C., \&
  {Young}, P.~R. 1997, \aaps, 125, 149

\bibitem[{{Ding} {et~al.}(2005){Ding}, {Li}, \& {Fang}}]{2005A&A...432..699D}
{Ding}, M.~D., {Li}, H., \& {Fang}, C. 2005, \aap, 432, 699

\bibitem[{{Fontenla} {et~al.}(2006){Fontenla}, {Avrett}, {Thuillier}, \&
  {Harder}}]{2006ApJ...639..441F}
{Fontenla}, J.~M., {Avrett}, E., {Thuillier}, G., \& {Harder}, J. 2006, \apj,
  639, 441

\bibitem[{{Goldberg}(1939)}]{1939ApJ....89..673G}
{Goldberg}, L. 1939, \apj, 89, 673

\bibitem[{{Golding} {et~al.}(2014){Golding}, {Carlsson}, \&
  {Leenaarts}}]{2014ApJ...784...30G}
{Golding}, T.~P., {Carlsson}, M., \& {Leenaarts}, J. 2014, \apj, 784, 30

\bibitem[{{Golding} {et~al.}(2016){Golding}, {Leenaarts}, \&
  {Carlsson}}]{2016ApJ...817..125G}
{Golding}, T.~P., {Leenaarts}, J., \& {Carlsson}, M. 2016, \apj, 817, 125

\bibitem[{{Gudiksen} {et~al.}(2011){Gudiksen}, {Carlsson}, {Hansteen}, {Hayek},
  {Leenaarts}, \& {Mart{\'{\i}}nez-Sykora}}]{2011A&A...531A.154G}
{Gudiksen}, B.~V., {Carlsson}, M., {Hansteen}, V.~H., {et~al.} 2011, \aap, 531,
  A154

\bibitem[{Gustafsson(1973)}]{gustafsson1973}
Gustafsson, B. 1973, Uppsala Astr.\ Obs.\ Ann., 5, No.\ 6

\bibitem[{{Hansteen}(1993)}]{1993ApJ...402..741H}
{Hansteen}, V. 1993, \apj, 402, 741

\bibitem[{{Ibgui} {et~al.}(2013){Ibgui}, {Huben{\'{y}}}, {Lanz}, \&
  {Stehl{\'e}}}]{2013A&A...549A.126I}
{Ibgui}, L., {Huben{\'{y}}}, I., {Lanz}, T., \& {Stehl{\'e}}, C. 2013, \aap,
  549, A126

\bibitem[{{Ji} {et~al.}(2012){Ji}, {Cao}, \& {Goode}}]{2012ApJ...750L..25J}
{Ji}, H., {Cao}, W., \& {Goode}, P.~R. 2012, \apjl, 750, L25

\bibitem[{{Jordan}(1975)}]{1975MNRAS.170..429J}
{Jordan}, C. 1975, \mnras, 170, 429

\bibitem[{{Joselyn} {et~al.}(1979){Joselyn}, {Munro}, \&
  {Holzer}}]{1979ApJS...40..793J}
{Joselyn}, J.~A., {Munro}, R.~H., \& {Holzer}, T.~E. 1979, \apjs, 40, 793

\bibitem[{{Kahler} {et~al.}(1983){Kahler}, {Davis}, \&
  {Harvey}}]{1983SoPh...87...47K}
{Kahler}, S.~W., {Davis}, J.~M., \& {Harvey}, J.~W. 1983, \solphys, 87, 47

\bibitem[{{Khomenko} \& {Collados}(2012)}]{2012ApJ...747...87K}
{Khomenko}, E. \& {Collados}, M. 2012, \apj, 747, 87

\bibitem[{{Kiselman} {et~al.}(2011){Kiselman}, {Pereira}, {Gustafsson},
  {Asplund}, {Mel{\'e}ndez}, \& {Langhans}}]{2011A&A...535A..14K}
{Kiselman}, D., {Pereira}, T.~M.~D., {Gustafsson}, B., {et~al.} 2011, \aap,
  535, A14

\bibitem[{{Landi} {et~al.}(2013){Landi}, {Young}, {Dere}, {Del Zanna}, \&
  {Mason}}]{2013ApJ...763...86L}
{Landi}, E., {Young}, P.~R., {Dere}, K.~P., {Del Zanna}, G., \& {Mason}, H.~E.
  2013, \apj, 763, 86

\bibitem[{{Leenaarts} \& {Carlsson}(2009)}]{2009ASPC..415...87L}
{Leenaarts}, J. \& {Carlsson}, M. 2009, in Astronomical Society of the Pacific
  Conference Series, Vol. 415, The Second Hinode Science Meeting: Beyond
  Discovery-Toward Understanding, ed. B.~{Lites}, M.~{Cheung}, T.~{Magara},
  J.~{Mariska}, \& K.~{Reeves}, 87

\bibitem[{{Leenaarts} {et~al.}(2007){Leenaarts}, {Carlsson}, {Hansteen}, \&
  {Rutten}}]{2007A&A...473..625L}
{Leenaarts}, J., {Carlsson}, M., {Hansteen}, V., \& {Rutten}, R.~J. 2007, \aap,
  473, 625

\bibitem[{{Leenaarts} {et~al.}(2012){Leenaarts}, {Carlsson}, \& {Rouppe van der
  Voort}}]{2012ApJ...749..136L}
{Leenaarts}, J., {Carlsson}, M., \& {Rouppe van der Voort}, L. 2012, \apj, 749,
  136

\bibitem[{{Leenaarts} {et~al.}(2013){Leenaarts}, {Pereira}, {Carlsson},
  {Uitenbroek}, \& {De Pontieu}}]{2013ApJ...772...90L}
{Leenaarts}, J., {Pereira}, T.~M.~D., {Carlsson}, M., {Uitenbroek}, H., \& {De
  Pontieu}, B. 2013, \apj, 772, 90

\bibitem[{{Lemen} {et~al.}(2012){Lemen}, {Title}, {Akin}, {Boerner}, {Chou},
  {Drake}, {Duncan}, {Edwards}, {Friedlaender}, {Heyman}, {Hurlburt}, {Katz},
  {Kushner}, {Levay}, {Lindgren}, {Mathur}, {McFeaters}, {Mitchell}, {Rehse},
  {Schrijver}, {Springer}, {Stern}, {Tarbell}, {Wuelser}, {Wolfson}, {Yanari},
  {Bookbinder}, {Cheimets}, {Caldwell}, {Deluca}, {Gates}, {Golub}, {Park},
  {Podgorski}, {Bush}, {Scherrer}, {Gummin}, {Smith}, {Auker}, {Jerram},
  {Pool}, {Soufli}, {Windt}, {Beardsley}, {Clapp}, {Lang}, \&
  {Waltham}}]{2012SoPh..275...17L}
{Lemen}, J.~R., {Title}, A.~M., {Akin}, D.~J., {et~al.} 2012, \solphys, 275, 17

\bibitem[{{Livshits} {et~al.}(1976){Livshits}, {Akimov}, {Belkina}, \&
  {Diatel}}]{1976SoPh...49..315L}
{Livshits}, M.~A., {Akimov}, L.~A., {Belkina}, I.~L., \& {Diatel}, N.~P. 1976,
  \solphys, 49, 315

\bibitem[{{MacPherson} \& {Jordan}(1999)}]{1999MNRAS.308..510M}
{MacPherson}, K.~P. \& {Jordan}, C. 1999, \mnras, 308, 510

\bibitem[{{Mart{\'{\i}}nez-Sykora} {et~al.}(2012){Mart{\'{\i}}nez-Sykora}, {De
  Pontieu}, \& {Hansteen}}]{2012ApJ...753..161M}
{Mart{\'{\i}}nez-Sykora}, J., {De Pontieu}, B., \& {Hansteen}, V. 2012, \apj,
  753, 161

\bibitem[{{Milkey} {et~al.}(1973){Milkey}, {Heasley}, \&
  {Beebe}}]{1973ApJ...186.1043M}
{Milkey}, R.~W., {Heasley}, J.~N., \& {Beebe}, H.~A. 1973, \apj, 186, 1043

\bibitem[{{Olluri} {et~al.}(2013{\natexlab{a}}){Olluri}, {Gudiksen}, \&
  {Hansteen}}]{2013ApJ...767...43O}
{Olluri}, K., {Gudiksen}, B.~V., \& {Hansteen}, V.~H. 2013{\natexlab{a}}, \apj,
  767, 43

\bibitem[{{Olluri} {et~al.}(2013{\natexlab{b}}){Olluri}, {Gudiksen}, \&
  {Hansteen}}]{2013AJ....145...72O}
{Olluri}, K., {Gudiksen}, B.~V., \& {Hansteen}, V.~H. 2013{\natexlab{b}}, \aj,
  145, 72

\bibitem[{{Olluri} {et~al.}(2015){Olluri}, {Gudiksen}, {Hansteen}, \& {De
  Pontieu}}]{2015ApJ...802....5O}
{Olluri}, K., {Gudiksen}, B.~V., {Hansteen}, V.~H., \& {De Pontieu}, B. 2015,
  \apj, 802, 5

\bibitem[{{Pereira} {et~al.}(2009){Pereira}, {Kiselman}, \&
  {Asplund}}]{2009A&A...507..417P}
{Pereira}, T.~M.~D., {Kiselman}, D., \& {Asplund}, M. 2009, \aap, 507, 417

\bibitem[{{Pesnell} {et~al.}(2012){Pesnell}, {Thompson}, \&
  {Chamberlin}}]{2012SoPh..275....3P}
{Pesnell}, W.~D., {Thompson}, B.~J., \& {Chamberlin}, P.~C. 2012, \solphys,
  275, 3

\bibitem[{{Pietarila} \& {Judge}(2004)}]{2004ApJ...606.1239P}
{Pietarila}, A. \& {Judge}, P.~G. 2004, \apj, 606, 1239

\bibitem[{{Pozhalova}(1988)}]{1988SvA....32..542P}
{Pozhalova}, Z.~A. 1988, \sovast, 32, 542

\bibitem[{{Rybicki} \& {Hummer}(1991)}]{1991A&A...245..171R}
{Rybicki}, G.~B. \& {Hummer}, D.~G. 1991, \aap, 245, 171

\bibitem[{{Rybicki} \& {Hummer}(1992)}]{1992A&A...262..209R}
{Rybicki}, G.~B. \& {Hummer}, D.~G. 1992, \aap, 262, 209

\bibitem[{{Sanz-Forcada} \& {Dupree}(2008)}]{2008A&A...488..715S}
{Sanz-Forcada}, J. \& {Dupree}, A.~K. 2008, \aap, 488, 715

\bibitem[{{Schad} {et~al.}(2013){Schad}, {Penn}, \&
  {Lin}}]{2013ApJ...768..111S}
{Schad}, T.~A., {Penn}, M.~J., \& {Lin}, H. 2013, \apj, 768, 111

\bibitem[{{Schad} {et~al.}(2015){Schad}, {Penn}, {Lin}, \&
  {Tritschler}}]{2015SoPh..290.1607S}
{Schad}, T.~A., {Penn}, M.~J., {Lin}, H., \& {Tritschler}, A. 2015, \solphys,
  290, 1607

\bibitem[{{Scharmer} {et~al.}(2003){Scharmer}, {Bjelksjo}, {Korhonen},
  {Lindberg}, \& {Petterson}}]{2003SPIE.4853..341S}
{Scharmer}, G.~B., {Bjelksjo}, K., {Korhonen}, T.~K., {Lindberg}, B., \&
  {Petterson}, B. 2003, in Society of Photo-Optical Instrumentation Engineers
  (SPIE) Conference Series, Vol. 4853, Innovative Telescopes and
  Instrumentation for Solar Astrophysics, ed. S.~L. {Keil} \& S.~V. {Avakyan},
  341--350

\bibitem[{{Smith}(2003)}]{2003MNRAS.341..143S}
{Smith}, G.~R. 2003, \mnras, 341, 143

\bibitem[{{Smith} \& {Jordan}(2002)}]{2002MNRAS.337..666S}
{Smith}, G.~R. \& {Jordan}, C. 2002, \mnras, 337, 666

\bibitem[{{Tobiska}(1991)}]{1991JATP...53.1005T}
{Tobiska}, W.~K. 1991, Journal of Atmospheric and Terrestrial Physics, 53, 1005

\bibitem[{{Tobiska}(2004)}]{2004AdSpR..34.1736T}
{Tobiska}, W.~K. 2004, Advances in Space Research, 34, 1736

\bibitem[{{Woods} {et~al.}(2012){Woods}, {Eparvier}, {Hock}, {Jones},
  {Woodraska}, {Judge}, {Didkovsky}, {Lean}, {Mariska}, {Warren}, {McMullin},
  {Chamberlin}, {Berthiaume}, {Bailey}, {Fuller-Rowell}, {Sojka}, {Tobiska}, \&
  {Viereck}}]{2012SoPh..275..115W}
{Woods}, T.~N., {Eparvier}, F.~G., {Hock}, R., {et~al.} 2012, \solphys, 275,
  115

\bibitem[{{Zarro} \& {Zirin}(1986)}]{1986ApJ...304..365Z}
{Zarro}, D.~M. \& {Zirin}, H. 1986, \apj, 304, 365

\bibitem[{{Zirin}(1975)}]{1975ApJ...199L..63Z}
{Zirin}, H. 1975, \apjl, 199, L63

\end{thebibliography}

 \bibliographystyle{aa} 

\end{document}